\documentclass{aa}
\usepackage[varg]{txfonts}
\usepackage{graphicx}
\usepackage{multirow}
\usepackage{tabularx}
\usepackage{array}
\usepackage{adjustbox}
\usepackage{linenoaa}
\usepackage{graphicx} 
\usepackage{subcaption} 
\newcolumntype{L}[1]{>{\raggedright\arraybackslash}p{#1}}
\usepackage{siunitx}
\usepackage{hyperref}
\usepackage{afterpage}  
\usepackage{placeins}   

\idline{692, A42 (2024)} 

\AANum{A42} 

\doi{10.1051/0004-6361/202449535} 

\hypersetup{
    colorlinks=true,
    linkcolor=blue,    
    citecolor=blue,    
    filecolor=blue,    
    urlcolor=blue      
}

\begin{document}


\title{The VANDELS Survey: Star formation and quenching in two over-densities at $3<z<4$}

\subtitle{}

\author{M. Espinoza Ortiz\inst{1,2}, L. Guaita\inst{2}, R. Demarco\inst{3}, A. Calabró\inst{4}, L. Pentericci\inst{4}, M. Castellano\inst{4}, M. Celeste Artale\inst{2}, N. P. Hathi\inst{5}, Anton M. Koekemoer\inst{5}, F. Mannucci\inst{4}, P. Hibon\inst{6}, D. J. McLeod\inst{7}, A. Gargiulo\inst{8} and E. Pompei\inst{6}}

\institute{Department of Astronomy, Universidad de Concepción, Casilla 160-C, Concepción, Chile
  \and Universidad Andres Bello, Facultad de Ciencias Exactas, Departamento de Fisica y Astronomia, Instituto de Astrofisica, Fernandez Concha 700, Las Condes, Santiago RM, Chile
    \and Institute of Astrophysics, Facultad de Ciencias Exactas, Universidad Andr\'es Bello, Sede Concepci\'on, Talcahuano, Chile
  \and INAF – Osservatorio Astronomico di Roma, Via di Frascati 33, 00078 Monte Porzio Catone, Italy
  \and Space Telescope Science Institute, 3700 San Martin Drive, Baltimore, MD 21218, USA
  \and European Southern Observatory, Santiago, Chile
  \and Institute for Astronomy, University of Edinburgh, Royal Observatory, Edinburgh EH9 3HJ, UK
  \and INAF – Istituto di Astroﬁsica Spaziale e Fisica cosmica Milano, Via Alfonso Corti 12, 20133 Milano, Italy}

\date{Received 7 February 2024 / Accepted 7 October 2024}

\authorrunning{Espinoza Ortiz et al.} 

\abstract {{Exploring galaxy evolution in dense environments, such as proto-clusters, is pivotal for understanding the mechanisms that drive star formation and the quenching of star formation. }
}
{{This study provides insights into how two over-densities could have impacted the physical properties, such as the star formation rate, stellar mass, morphology, and the evolution of their members, particularly members characterised by a quenching of star formation.}
}
{{We focus on the over-densities identified at $3<z<4$ in the Chandra Deep Field South (CDFS) and in the Ultra Deep Survey (UDS) regions of the VIMOS (VIsible MultiObject Spectrograph) Ultra Deep Survey (VANDELS). Our methodology involves the analysis of the spectral energy distribution of the members of the over-densities and of the galaxies in the field. We relied on Bayesian analysis techniques BEAGLE and BAGPIPES to study the best-fit physical parameters and the rest-frame U-V and V-J colours (UVJ). This approach allowed us to separate quenched and star-forming galaxies based on the UVJ diagram and by estimating their specific star formation rate (sSFR). We used the TNG300 simulation to interpret our results.}
}
{{We find that two out of 13 over-densities host quenched galaxies, with red rest-frame U-V colour and low sSFR. The physical properties of them are consistent with those of massive passive galaxies from the literature. The quenched members are redder, older, more massive, and show a more compact morphology than the other galaxy members. The two over-densities, with the highest-density peaks at $z\simeq3.55$ and $z\simeq3.43,$ respectively, have dark matter halo masses consistent with being proto-clusters at $z\sim3$ and they each host an active galactic nucleus (AGN). We found five AGNs in the structure at $z\simeq3.55$ and three AGNs in the one at $z\simeq3.43$. In comparison to quenched galaxies in the field, our massive quenched members show a higher local density environment.  By using the IllustrisTNG simulation (TNG300), we find that proto-cluster structures with quenched galaxies at high redshift are likely to evolve into a structure with a higher fraction of passive galaxies by $z = 1$.
}}
{{The two over-densities studied here host massive quenched galaxies in their highest-density peaks and AGNs. By following the evolution of the passive galaxies in the simulated proto-clusters at $z = 3$ from the TNG300 simulation, we find that the median of their sSFRs was larger than 10$^{-8}$ yr$^{-1}$ at $z=6$ and the median mass growth rate was  96\% from $z=6$ to $z=3$. In 20\% of the simulated proto-clusters, the passive galaxy had already accreted 10-20\% of the mass at $z=6$, with SFRs $>100$ M$_{\odot}$ yr$^{-1}$ at $z=8$. The conditions for this favorable mass assembly could be the galaxy interactions and the high gas accretion rate in the dense environment. As a consequence, the quenching of the star formation at $z=3$ could be driven by the black hole mass growth and AGN feedback. This scenario is consistent with the properties of the two quenched galaxies we find in our two over-densities at $z\sim3$.}}


\maketitle

\nolinenumbers
\section{Introduction}

The investigation into how galaxies evolve within proto-clusters and dense cosmic structures has been a pivotal area of astrophysical research, elucidating key mechanisms of galaxy formation, interaction, and quenching processes \citep{gunn72, dressler97, balogh99, ebeling01, weinmann06, olsen07, wetzel12, menanteau13, wetzel13}. These studies have significantly advanced our understanding by highlighting the influence of environmental factors on galaxy morphology, star formation rates, and the role of active galactic nuclei (AGNs) in shaping galaxy evolution across cosmic time.\\
\\
{Galaxies are expected to experience early growth at an accelerated rate in over-densities (regions where the galaxy density is significantly higher than the average) at high redshift \citep[e.g.][]{chiang13}, where the intensity of star formation is expected to surpass that of the field.} Observations at $z>2.5$ have reported higher star formation (SF) in over-density galaxies compared to field galaxies (e.g. \citealp{lemaux22} and \citealp{Darvish2016}). Some of the processes that may be driving this rapid star formation include galaxy interactions, mergers { \citep{Lotz2011}, and gas accretion. The} significance of cold gas accretion as a mechanism for star formation in early galaxies within over-densities has been highlighted by \citet{Dekel2009} and \citet{Keres2005}.\\
\\
Based on simulations, \citet{chiang17} showed that the contribution of the galaxies in over-densities to the cosmic star formation rate density rises from 20\% at $z=2$ to 50\% at $z=10$. This increase underscores the critical role of over-densities in the cosmic star formation history and is supported by other simulation works, such as those of \citet{overzier09} and \citet{pillepich18}. These studies have further detailed the mechanisms by which over-densities contribute to the broader context of galaxy formation and evolution.\\
\\
{In the Local Universe, galaxies in clusters have evolved and exist in a passive state, where star formation has been stopped by phenomena that could be related to the environment} \citep{Baldry06,boselli06,Peng2010}. {More evolved galaxies are typically redder and more massive, so we expect to find such galaxies with those physical properties more often in clusters than in the field.}\\
\\
The mechanisms that drive the increase and quenching of star formation in galaxy clusters are diverse and complex. 
Galaxy interactions \citep[e.g.][]{park09} play a significant role in both enhancing and quenching star formation. Starvation, a process that prevents the inflow of gas essential for star formation, can often be caused by intra-cluster gas heating \citep{shahidi20, feldman15}. Ram pressure stripping, where the gas is removed from galaxies as they move through the intra-cluster medium, further inhibits star formation by directly stripping away the galaxy's gas reservoir \citep{gunn72, Abadi1999}.\\
\\
Galaxy mergers (e.g. \citealp{pontzen16,davies22}) can also play a significant role in galaxy evolution, often triggering star formation and structural transformations. Environmental interactions, such as tidal interactions and harassment (e.g. \citealp{tal14}), ram-pressure stripping of the gas (e.g. \citealp{peng19}), and the quenching of star formation through starvation when the hot intra-cluster medium cuts off the gas supply to galaxies (e.g. \citealp{foltz18,mao22}), significantly affect galaxy properties. These processes can lead to the removal of gas, halt star formation, and transform the galaxy's morphology, underscoring the complex interplay between galaxies and their surroundings.\\
\\
{AGN feedback that is not necessarily related to environment is another important mechanism to consider. Such feedback is the result of  AGNs releasing energy in the surrounding medium. A consequence of this is that the surrounding gas is heated, subsequently inhibiting cooling and star formation, and it is also expelled from the galaxy} \citep{fabian12, schawinski07}. Other processes contributing to the quenching of star formation include gravitational heating \citep{johansson09, dekel08}, stellar feedback mechanisms \citep{hopkins12, collins22}, and morphological changes within galaxies \citep{Lu22, kim18}. \citet{peng19} introduced the concept of 'morphological quenching', where the growth of a central bulge in a galaxy impedes star formation in the surrounding disk. These authors highlighted the role of halo quenching, where the heating of baryons in the circumgalactic medium occurs in halos above a certain mass threshold.\\
\\
{
Recent studies \citep[e.g.][]{shi21,mcconachie22,shi24} have found star-forming galaxies that are simultaneously evolved  with quenched star formation at $z>2$. By studying these sources, we can investigate the processes that contribute to accelerate star formation and to its halt at their high redshifts. 
}\\
\\
{Proto-clusters}, the precursors of local galaxy clusters \citep{chiang13}, are of particular interest in the study of the phenomena that can lead to accelerated episodes of SF and then its quenching. They provide a unique window into the early stages of large-scale structure assembly, evolution, and quenching of galaxies in them. By examining these proto-clusters, we can gain insight into the initial processes that lead to the formation of passive galaxy populations.\\
\\
In this work, we delve into the over-densities detected in \citet{guaita2020} within the Chandra Deep Field South (CDFS; \citealp{Giacconi2002}) and Ultra Deep Survey (UDS; \citealp{Lawrence2007}) at an epoch of \(3 < z < 4\). This epoch marks a critical phase in the early history of mass assembly in the universe. {Our primary goal is to analyse the galaxies residing in these dense regions to unravel both the external and internal factors influencing their evolution.  
}
{This 
investigation  offers a look into the dynamics at play in the 
early stages of cosmic structure formation.}\\
\\
The paper is organized as follows. In Sect. \ref{sec:data}, we present our data. In Sect. \ref{sec:methodology}, we explain the process of  estimating the colours and the selection criteria for passive galaxies, along with a comparison of passive galaxies in the literature. In Sect. \ref{sec_result}, we show the scaling relations. In Sect. \ref{sec:discution}, we study  the over-densities with passive galaxies in detail. Throughout this paper, we use an AB magnitude \citep{OkeGunn1983} and we assume a cosmology with \( H_0 = 70 \, \text{km s}^{-1} \text{Mpc}^{-1} \), \( \Omega_m = 0.3 \), and \( \Omega_\Lambda = 0.7 \).\\
\\
\section{Data}
\label{sec:data}
The data used in this work come from the deep VIMOS survey of the CANDELS CDFS and UDS fields (VANDELS; \citealp{vandels}). We took advantage of the comprehensive photometric catalog (\citealp{pente18}, \citealp{mcclure18}, \citealp{garili21} and \citealp{talia23}), which includes photometric redshifts and a variety of physical properties, such as stellar mass, star formation rates (SFRss), mass-weighted age, star-formation time scale ($\tau$), and specific SFR (sSFR).\\
\\
Our study utilises the over-dense structures identified in \citet{guaita2020} within the VANDELS fields. We have updated the redshifts of the over-density members and field galaxies to the most recent spectroscopic redshifts from VANDELS. We also considered the most updated compilation of redshift from the literature made by Nimish Hathi (priv. comm.) for the VANDELS collaboration.

\subsection{Spectral energy distribution analysis}

{The physical properties in the VANDELS photometric catalog were primarily estimated using the BayEsian Analysis of GaLaxy sEds (BEAGLE) tool \citep{chevellard16}. In addition, for further validation, we employed the Bayesian Analysis of Galaxies for Physical Inference and Parameter EStimation (BAGPIPES) software \citep{carnall_bagpipes}, which is particularly well-suited for dusty star-forming and quiescent galaxies (see Sect. \ref{sec:passives}).}\\
\\
{The configuration details for the fitting of the spectral energy distribution (SED) performed with BEAGLE are described in \citet{calabro22} and \citet{castellano23}. BEAGLE incorporates both stellar and nebular emissions coming from a galaxy, without including an eventual active galactic nucleus (AGN) component. Nebular emission is represented following the \citet{gutkin16} approach with a Chabrier initial mass function \citep{chabrier}. The star-formation history (SFH) is represented by an exponentially delayed form, 
\begin{equation}
 SFR(t) \; \propto \; t \cdot \; e^{({{-t}/{\tau}})},   
\end{equation}
where $t$ is the age of the galaxy and $\tau$ the e-folding time of the star-forming period. 
When $\tau>t$, the galaxy is observed during the star-forming period. When $\tau<t$, the galaxy is observed after the peak of the main episode of star formation.}\\
\\
{To further validate the physical properties of a subset of galaxies in our sample, we used the BAGPIPES software, which is particularly effective for modelling the SEDs of dusty star-forming and quiescent galaxies \citep{Carnall2020}. The BAGPIPES analysis was conducted using the same delayed star-formation history as BEAGLE, with metallicity, mass-weighted age, stellar mass, tau e-folding time, and dust reddening as free parameters.}\\
\\
We also made use of images from the \textit{Hubble} Space Telescope (HST), including the high-level science mosaics from CANDELS \citep{Koekemoer2011, Grogin2011},  to examine the morphological properties of the objects in our samples {and relate them to their physical properties.} 

\subsection{Proto-cluster identification and member confirmation}
\begin{figure}[!htbp]
    \resizebox{\hsize}{!}
    {\includegraphics{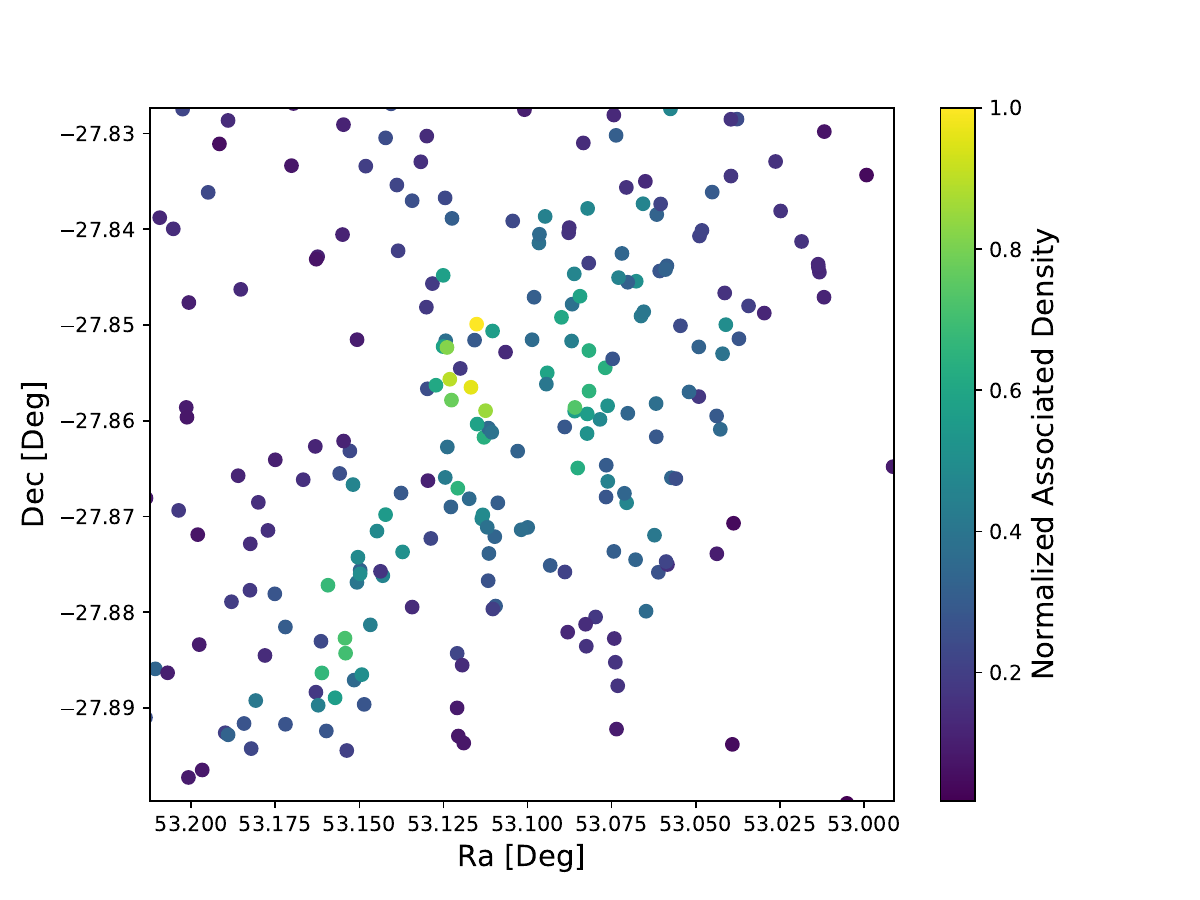}}
    \caption{Spatial distribution and associated density of the galaxies associated with the \(z355\) overdensity (see Table 1) and the corresponding field galaxies. The X-axis represents the RA and the Y-axis the Dec in degrees. colours indicate the associated density, normalized with respect to the highest density peak observed in \(z355\).}
    \label{fig:RA_DEC_density}
\end{figure}

The over-densities studied in this work were originally selected in \citet{guaita2020}. We summarize here the main criteria used in the selection and refer to that paper for more details. The estimation of the number density of galaxies was based on the algorithm developed by \citet{trevese07, Salimbeni2009}. This algorithm considers as an input the 3D coordinates of all the objects in the catalog, namely the right ascension (RA), declination (Dec) in the sky, and the redshift. {We used the spectroscopic redshift when available, but for the majority of the objects, we relied on photometric redshifts. The entire survey volume was initially divided into 3D cells in the RA, Dec, and redshift dimensions. As explained in \citet{guaita2020}, the initial volume of the cells depends on the spatial accuracy and on the photometric redshift uncertainty ($0.02 \times$ (1 + $z$) for the VANDELS catalog).}\\
\\
Then, every cell was increased in volume till a number of 10 galaxies was reached. Therefore, the local density of every galaxy contained in one of the cells of the original grid was defined as 10 divided by the final volume of the cell. The algorithm identified galaxy over-densities by comparing the local density value in each volume cell to the mean density at a given redshift \citep[see also][]{calabro22}. The over-density centres were defined as the density peaks of the highest signal to noise. An example of one of the over-densities is shown in Fig. \ref{fig:RA_DEC_density}.
The galaxies with redshifts within the redshift range of a certain over-density and spatially located outside the over-density region compose the field.\\
\\
The extensive coverage and depth of the VANDELS survey in the CDFS and UDS fields, coupled with high levels of completeness in the photometric catalogs, enable us to construct a robust and comprehensive redshift framework for our study. Therefore, we verified the over-density members double-checking their redshift with the most updated spectroscopic catalog provided to the VANDELS collaboration \citep[Nimish Hathi's spectroscopic redshift compilation and VANDELS DR4][]{garili21}. The identified over-densities together with their number of spectroscopic and photometric redshifts are listed in Table \ref{tab:protoclusters_data}. The histograms of the redshift distributions of all the over-densities available on Zenodo (see Sec. \ref{data_av}).

\subsection{AGN identification}
{To identify the AGN hosted by our over-densities, we utilised X-ray catalogs from the literature, specifically the catalogs of \citet{koce18} for UDS and \citet{chandra} for CDFS.} 
We also cross-referenced these catalogs with the VANDELS catalog of emission-line AGNs (Bongiorno A. et al., private communication). {We used a positional matching criterion within 1 arcsec to ensure high accuracy in our AGN identification process. This 
approach led to the identification of eight AGNs within our CDFS over-densities.}\\
\\
{For the sources identified as AGN, we cannot trust the physical parameters obtained through SED fitting, since the BEAGLE and BAGPIPES runs did not include AGN templates. Therefore, we acknowledge the presence of AGN, but we do not further discuss their properties in this work.}
%

\subsection{Morphological analysis}
We used the catalog from \citet{vander12} to associate a half-light radius to the members of the over-densities and to the field galaxies. The radius was calculated using the photometry in the F160W filter and adopting the GALFIT (\citealp{Peng2002,Peng2010}) algorithm on the \textit{Hubble} Space Telescope F160W image. {The catalog includes a flag from GALFIT, which indicates the quality of the fit. We accepted measurements only when the GALFIT flag is equal to 0, signifying a good fit. In our analysis, we found that the entire sample of $3302$ galaxies has a good flag, with 190 being members and $3112$ being field galaxies.}\\

\section{Methodology}\label{sec:methodology}

\subsection{Identification of passive galaxies}\label{sec:passives}
\begin{figure}[!htbp]
    \resizebox{\hsize}{!}{\includegraphics{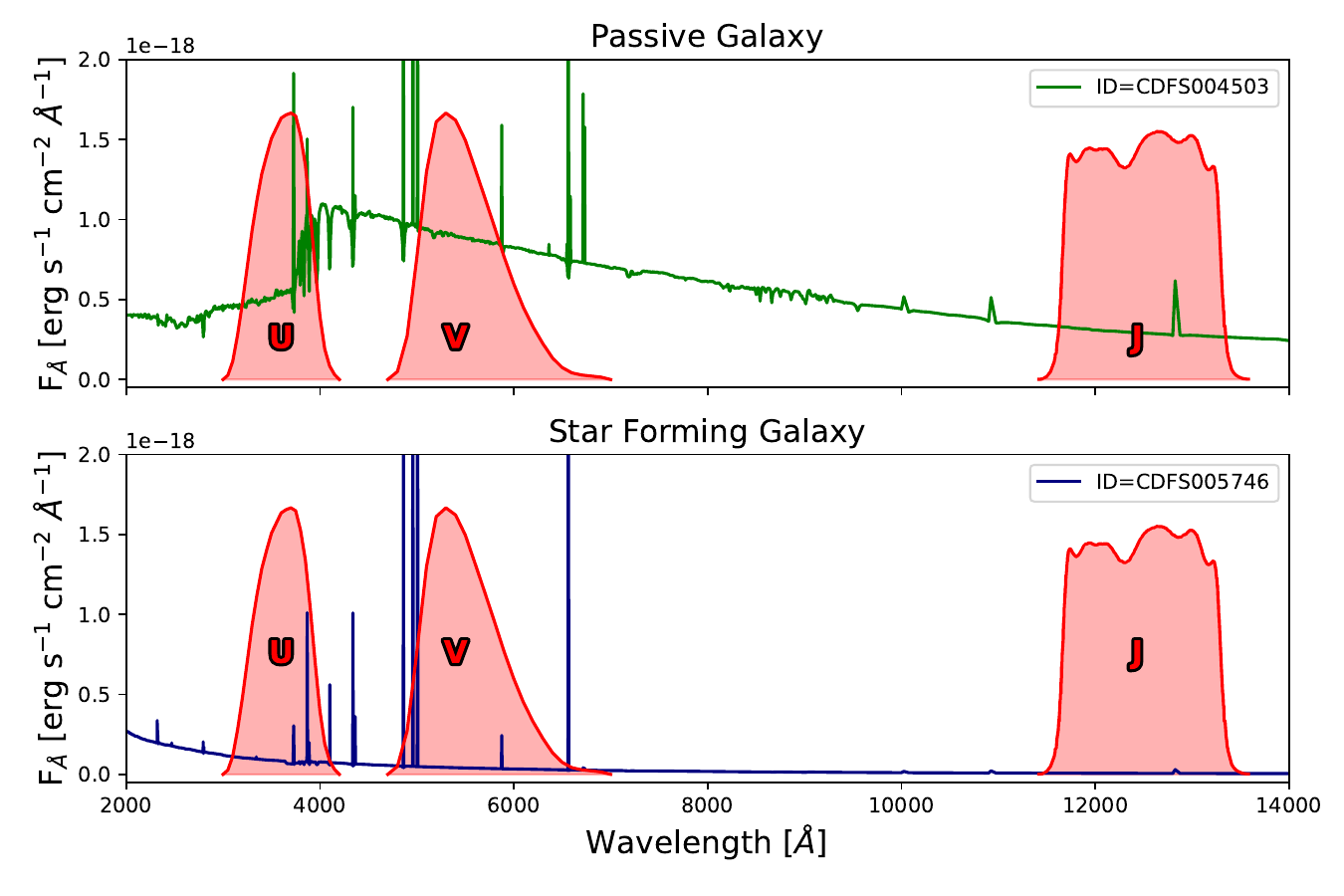}}
    \caption{Comparison of SEDs computed by BEAGLE for two galaxy types. The upper panel displays the SED of a passive galaxy (green line), whereas the lower panel shows the SED of a star-forming galaxy (navy line). {The U (3584 \AA), V (5477 \AA), and J (12474 \AA) filters}, utilised for calculating photometric magnitudes, are marked on each graph to highlight the spectral regions where these galaxies exhibit significant differences in the rest-frame.}
    \label{fig:SEDs}
\end{figure}

{We constructed the UVJ diagram (rest frame U-V versus rest frame V-J colours) to distinguish star-forming from quiescent galaxies \citep[e.g.][]{carnall18, merlin19}. This is important to understand the evolutionary state of the members of our identified over-densities.
As depicted in Fig. \ref{fig:SEDs}, the SEDs of star-forming and quiescent galaxies appear different in the wavelength range of the rest-frame U, V, and J filters}. Quiescent or passive galaxies 
show higher flux at longer wavelengths, indicative of redder spectra. In contrast, star-forming galaxies show higher flux at shorter wavelengths or bluer spectra. For the examples in Fig. \ref{fig:SEDs}, the passive galaxy (upper panel) presents rest-frame $U-V=1.34$ and rest frame $V-J=0.61$, while the star-forming galaxy (lower panel) presents rest-frame $U-V=0.17$ and rest-frame $V-J=-0.21$.\\
\\
{To 
calculate the rest-frame U, V, and J magnitudes, we followed the methodology proposed by \citet{williams09}, employing 
their filter transmission curves and defining as passive the galaxies with rest-frame $U-V>1.3$ and $V-J \lesssim 1$ (Fig. 3).
We selected the most representative SED model 
for each galaxy, we translated it into the rest frame, and we convolved with the U, V, and J filter transmission curves. 
As the most representative SED model, we chose the one provided by BEAGLE that corresponds to the best-fit 
}
physical parameters, such as stellar mass, star-formation rate, star-formation history timescale, $\tau$, mass-weighted age, and dust attenuation.\\ 
\\
By following the above-mentioned criterion, the categorization of a galaxy as 'star-forming' hinges on its location outside the designated quiescent region of the UVJ diagram (see Sec. \ref{data_av}).\\
\\
%
{In order to corroborate if the selected galaxies were quiescent, we checked their sSFR values. \citet{carnall18} proposed that for quiescent galaxies their 
{instantaneous sSFR} should be less than 0.2 divided by the age of the Universe at the redshift of the galaxy ($t_{obs}$). Both the UVJ diagram and the sSFR conditions 
were used and we selected a total of 10 passive-galaxy candidates.\\
\\
To ensure the accuracy of our selection, we performed an additional analysis using the BAGPIPES software \citep{carnall_bagpipes}, specifically to verify the sSFR of these candidates. This further analysis confirmed the low sSFR for two out of the ten passive galaxy candidates. The BAGPIPES fit provided parameters consistent with those of dusty star-forming galaxies for the other eight candidates.\\
\\
Therefore, we excluded eight candidates from our final analysis, retaining only the two passive galaxies that met both the sSFR criterion from BAGPIPES and the initial BEAGLE criteria. It is important to note that we used BAGPIPES only to check the sSFR and did not re-evaluate the UVJ classification. Thus, the final selection consisted of galaxies satisfying the following cuts: 1) UVJ passive from BEAGLE, 2) sSFR $< 0.2/t_{obs}$ from BEAGLE, and 3) sSFR $< 0.2/t_{obs}$ from BAGPIPES.}

\subsection{Passive galaxies from literature}
We conducted a cross-matching procedure with several catalogs of passive galaxies from the literature, specifically, \citet{merlin19}, \citet{schreiber18}, \citet{shahidi20}, and \citet{straat14}. {This cross-matching between the entire VANDELS catalog and the listed publications resulted} in the identification of ten sources, including over-density members and galaxies in the field, that were previously recognized in the literature. Three of the ten cross-matches are in the field of the over-density at $z = 3.23$ (see Table \ref{tab:protoclusters_data}). However, for the CDFS016526 galaxy (nomenclature from the VANDELS catalog) BEAGLE did not converge and so did not provide a meaningful fit and reliable physical parameters, while for the CDFS013394 and CDFS019446 galaxies, BEAGLE provided UVJ colours and sSFRs inconsistent with typical values of passive galaxies. {Another} five of the ten cross-matches are members of the over-density at $z = 3.55$ (see Table \ref{tab:protoclusters_data}): {the CDFS004503 galaxy  is consistent with being passive, while CDFS005479 and CDFS006294 host AGNs. The CDFS003718 and CDFS004587 galaxies have UVJ colours and sSFRs inconsistent with typical values of passive galaxies. Two of the ten crossmatched galaxies are members of the over-density at \( z = 3.43 \) (see Table \ref{tab:protoclusters_data}): CDFS019505 hosts an AGN, and CDFS019883 is consistent with being passive.}\\
\\
In the analysis of the over-density at \( z = 3.55 \) (see Sect. \ref{sub:z355}), we will consider one of our confirmed passive galaxies and also two passive galaxies from the literature, CDFS004587 and CDFS003718. These galaxies were identified as passive in previous studies but were not selected as such in our data. \\
\\
The CDFS004503 and CDFS019883 passive galaxies were identified in both \citet{merlin19} and \citet{straat14}. \citet{merlin19} 
utilised ALMA and \textit{Spitzer} data to specifically avoid the misidentification of dusty starbursts as passive galaxies. Similarly, \citet{straat14} employed \textit{Spitzer}/MIPS 24 $\mu$m 
and the \textit{Herschel}/PACS 100 $\mu$m and 160 $\mu$m observations for the same purpose, ensuring that these galaxies are indeed confirmed as passive -- and not mistaken as dusty starbursts. 
\\
{Our analysis revealed that a subset of the passive galaxies identified in the literature are located within the over-densities, we study here. This finding is significant because it highlights that 
environmental factors may play a critical role in  galaxy evolution and in the halt of the star formation. 
While the number of objects involved is relatively small, the relation between passive galaxies and dense regions offers valuable insights into galaxy evolution processes that warrant further investigation.}\\
\\
%
%
Further discussion on the implications of these findings and a more detailed comparison between the literature-based and our identified passive galaxies will be presented in the Sect. \ref{sec:discution}. This will include an examination of the spatial distribution of the passive galaxies within the over-densities and an analysis of their potential role in the 
evolution of these dense environments.

\section{Results}\label{sec_result}
\subsection{Properties of the over-densities at $3<z<4$}
\begin{figure*}[!htbp]
    \resizebox{\hsize}{!}{\includegraphics{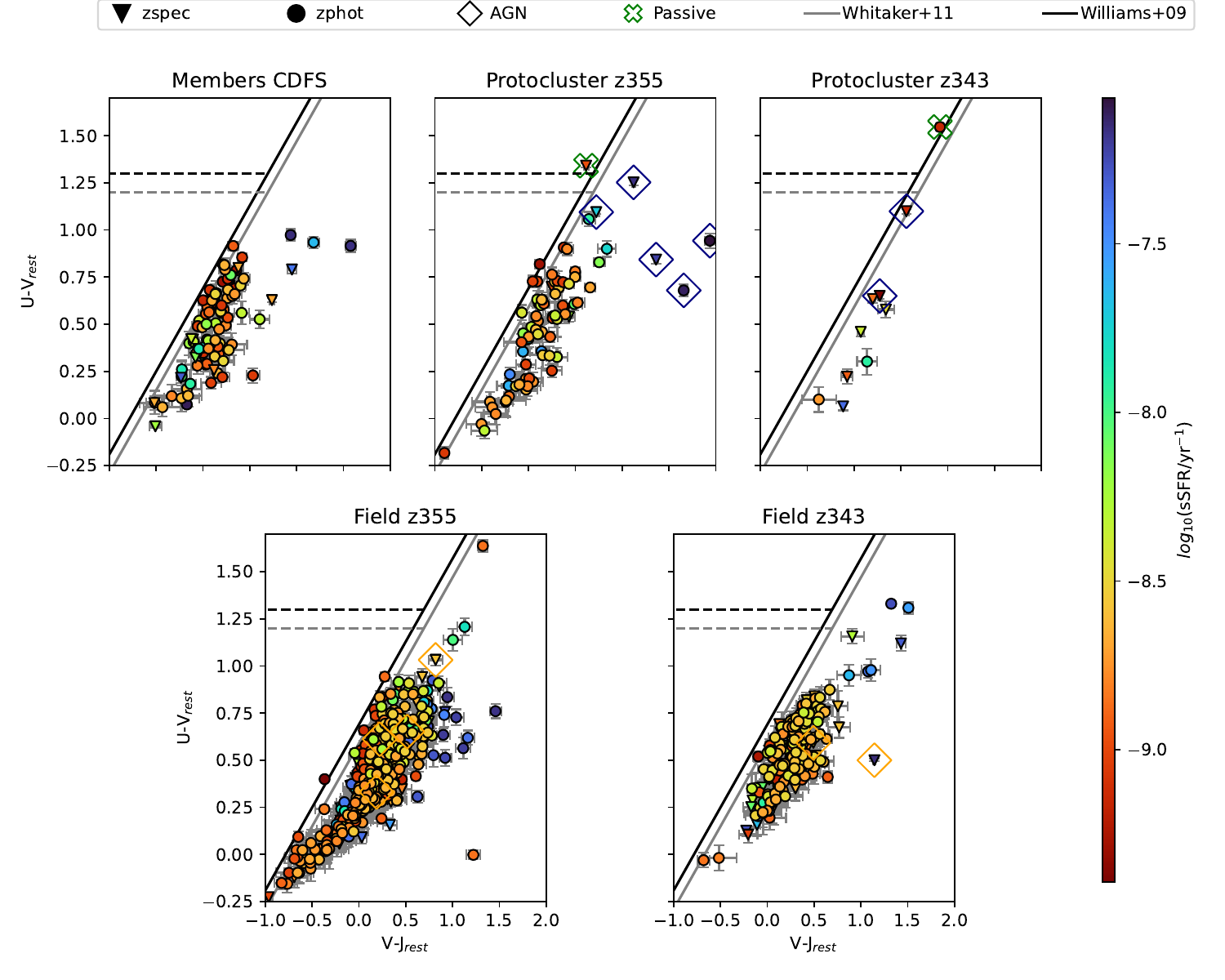}}
    \caption{{Rest-frame U-V versus V-J diagrams showcasing galaxies in different environments
    Upper left: galaxies that are members of the over-densities identified in CDFS excluding the two that host passive galaxies. Upper centre: members of the proto-cluster at $z=3.55$. Top-right: members of the proto-cluster at $z=3.43$. Lower left: field galaxies associated with the proto-cluster at $z=3.55$. Bottom-right: field galaxies associated with the proto-cluster at $z=3.43$. The colour scale denotes log(sSFR) provided by BEAGLE, with red signifying lower values and blue higher ones. Different symbols 
    represent 
    galaxies with a spectroscopic redshift (triangles), galaxies with a photometric redshift (dots), 
    AGNs (diamonds), and 
    passive galaxies (creen crosses). The lines in the diagrams outline the classification criteria set by \citet{williams09}, we used for the passive galaxy selection, and \citet{whitaker11}, shown for comparison.}}
    \label{fig:uvj_all}
\end{figure*}

{For all the over-densities studied here, we report in Table \ref{tab:protoclusters_data} the coordinates of their highest density peak, an identification name that will be used throughout the paper, their maximum volume, and their estimated dark matter halo mass.} 
The first six over-densities were identified in CDFS and the other seven 
in UDS. As a volume, we report the cylindrical volume implied by the redshift distribution, which is affected by photometric redshift errors. As dark matter halo mass, we report the value obtained from the total stellar mass of the members and the calibration from \citet{burg14}.\\
\begin{table*}[!htbp]
\tiny
\caption{Characteristics of the over-densities studied in this work.}
\label{tab:protoclusters_data}
\centering
\begin{tabular}{
    c c c c c 
    S[table-format=6.0] @{\,$\pm$\,} S[table-format=3.0] 
    @{\vspace{+0.1cm}\kern+2ex}S[table-format=4.2] 
    @{\kern+1ex\,[\kern+1.5ex} S[table-format=1.2]  
    @{,\,\kern+1ex} S[table-format=0.3] @{\kern-1ex\,]\kern+3ex}  
    c c c
}
\hline\hline
\begin{tabular}[c]{@{}c@{}}RA\\ {[}J2000{]}\end{tabular} & \begin{tabular}[c]{@{}c@{}}DEC\\ {[}J2000{]}\end{tabular} & Field & Name & N & \multicolumn{2}{c}{\begin{tabular}[c]{@{}c@{}}Volume\\ {[}Mpc$^3${]}\end{tabular}} & \multicolumn{3}{c}{\begin{tabular}[c]{@{}c@{}}Dark Matter Halo\\ {[}10$^{12}$ M$_\odot${]}\end{tabular}} & AGN & zspec & zphot  \\  
\hline
\\[-5pt]
53.1346 & -27.6954 & CDFS & z323 & 27 & 8186 & 614 & 0.74 & 2.33 & 0.13 & 0 & 3 & 24 \\ 
53.1229 & -27.7404 & CDFS & z329 & 31 & 13356 & 581 & 3.60 & 8.69 & 0.93 & 0 & 3 & 28 \\
53.0137 & -27.7388 & CDFS & z343 & 10 & 281 & 91 & 14.90 & 28.60 & 5.44 & 2 & 7 & 3 \\
53.0985 & -27.8060 & CDFS & z354 & 22 & 5645 & 871 & 0.49 & 1.63 & 0.07 & 0 & 4 & 18 \\
53.1187 & -27.8596 & CDFS & z355 & 82 & 31230 & 1974 & 28.50 & 49.30 & 12.10 & 5 & 8 & 74 \\
53.0712 & -27.6921 & CDFS & z369 & 18 & 3184 & 445 & 0.54 & 1.79 & 0.09 & 0 & 5 & 13 \\
34.2602 & -5.2498 & UDS & z324 & 31 & 5527 & 829 & 3.17 & 7.81 & 0.79 & 0 & 4 & 27 \\
34.4552 & -5.2023 & UDS & z327 & 20 & 1422 & 266 & 1.32 & 3.76 & 0.27 & 0 & 1 & 19 \\
34.5203 & -5.1648 & UDS & z332 & 29 & 8015 & 903 & 0.69 & 2.17 & 0.12 & 0 & 4 & 25 \\
34.4027 & -5.1648 & UDS & z349 & 7 & 139 & 34 & 0.13 & 0.55 & 0.01 & 0 & 1 & 6 \\
34.3427 & -5.2423 & UDS & z351 & 10 & 1230 & 131 & 0.17 & 0.67 & 0.21 & 0 & 0 & 10 \\
34.5186 & -5.2356 & UDS & z365 & 11 & 994 & 164 & 0.29 & 1.07 & 0.42 & 0 & 4 & 7 \\
34.5427 & -5.2023 & UDS & z369 & 48 & 18090 & 1940 & 0.78 & 16.70 & 2.45 & 0 & 5 & 43 \\
53.0705 & -27.8685 & CDFS & G-A & 39 & 9938 & 652 & 3.09 & 7.66 & 0.77 & 2 & 5 & 34\\
53.1132 & -27.8698 & CDFS & G-B & 25 & 2915 & 188 & 13.60 & 26.50 & 4.87 & 3 & 3 & 22 \\
53.1593 & -27.8771 & CDFS & G-C & 16 & 1803 & 164 & 0.84 & 2.57 & 0.15 & 0 & 0 & 16 \\
\hline
\end{tabular}\\
\medskip 
    \tiny 
    {Note.} The table columns correspond to (1) right ascension and (2) declination of the highest density peak, (3) field, (4) nomenclature, (5) total number of members, (6) maximum volume given by the redshift range of the members, the uncertainty corresponds to the photometric redshift uncertainty of the VANDELS catalog, (7) dark matter halo mass, estimated from the total stellar mass of members using the calibration provided by \citet{burg14}. Values in brackets indicate the lower and upper limits according to this calibration, (8) number of X-ray-detected and emission line AGN, (9) number of spectroscopic redshifts, and (10) number of photometric redshifts for the over-density members.
\end{table*}\\
{In our study of the 13 identified over-densities, 
we found that only two galaxies fulfilled the passive criteria. They are hosted by the 
\(z355\) and \(z343\) over-densities, as highlighted in Fig. \ref{fig:uvj_all}.} 
To enrich our analysis and provide a comparative perspective on the characteristics of passive galaxies across different environments, we considered 
a 
passive galaxy from the \(z323\) field (CDFS013394). This decision aims to juxtapose the properties of passive galaxies found in the dense environments of over-densities against those situated in the less dense, field regions. Notably, the proto-cluster environments harboring passive galaxies were also found to contain AGNss. This was not 
observed in the UDS over-densities, which lack both passive galaxies and AGN.\\
\\
It is important to clarify that in Sect. \ref{sec_result}, 
the focus is narrowly placed on the properties and spatial distribution of passive galaxies within these dense over-densities, without incorporating the passive galaxy from the field. This approach allows for a detailed examination of the unique conditions and evolutionary states of the two passive galaxies in over-densities.\\
\\
\subsection{Scaling relations}
{In Fig. \ref{fig:SFR_Mass}, we present 
the star-formation rate versus stellar mass relation 
for the galaxies in our sample. It shows 
a distinct 
trend} 
between ongoing star formation processes and galaxy stellar mass 
\citep{santini17}. 
{
We also show the rest-frame U-V colour versus stellar mass in Fig. \ref{fig:uv_Mass}. 
The galaxies identified as passive in our analysis are notably the reddest and most massive among the members, 
also compared to the members of the 
other proto-clusters. Specifically, the rest-frame U-V colours of the passive galaxies are larger than 1.3 by definition and most of the other galaxies typically have values lower than 1 (except for the few dusty star-forming galaxies which are also characterised by red rest-frame V-J colours). The stellar masses provided by BEAGLE for the passive galaxies are larger than $10^{11} \, M_{\odot}$, while the other members and the field galaxies have masses of the order of $10^8 - 10^{11} \, M_{\odot}$. 
} 
{The fact that passive galaxies are more massive and display redder colours than star-forming galaxies 
is also consistent with theoretical expectations \citep{Bell2004, Blanton2009, Peng2010} and implies specific characteristics of the stellar populations. }\\
\\
In Fig. \ref{fig:SigmaSFR_Mass}, we present the relationship between the SFR surface density ($\Sigma_{\text{SFR}}$) and stellar mass \citep{Salim2023, Calabro2024}. The $\Sigma_{\text{SFR}}$ is here calculated as the instantaneous SFR divided by the galaxy's area, which is determined using the formula $2\pi r^2$, where $r$ is the galaxy's half-light radius (Sect. 2.4). 
{By drawing a diagonal line as a trend for the $\Sigma_{\text{SFR}}$ versus stellar mass of star-forming galaxies at $3<z<4$, 
passive galaxies tend to exhibit a lower $\Sigma_{\text{SFR}}$ 
at their stellar mass.
As proposed in \citet{Salim2023}, the $\Sigma_{\text{SFR}}$ versus stellar mass together with the sSFR versus stellar mass diagram provide a valuable evolutionary perspective, revealing the level of star formation activity in galaxies. \citet{Salim2023} also emphasized that $\Sigma_{\text{SFR}}$, by being tied to molecular gas density and stellar feedback effectiveness, offers a clear distinction between different galaxy types, such as starburst and spheroid-dominated galaxies. Galaxies with globally low star-formation activity would be characterised by both low sSFR and low $\Sigma_{\text{SFR}}$ at their stellar mass. 
The values of $\Sigma_{\text{SFR}}$ that we calculated for the passive galaxies in our sample 
support the definition of their passive nature.} \\
{To study the size of our passive galaxies, we analysed their half light radius as a function of stellar mass (Fig. \ref{fig:Size_Mass}). We observed that the passive galaxies are compact and have stellar masses and sizes consistent with those of early-type galaxies at $z\sim3$ 
\citep{vander14}\\ 
\\
The majority of the other over-density members and of field galaxies have instead half-light radii typical of late-type galaxies. 
\\
The compactness of passive galaxies carries significant implications. 
Considering 
the fact that these galaxies are also the most massive, we can infer that they possess a high stellar density. In other words, the passive galaxies in our sample are not only massive, but their stellar mass is concentrated in a relatively small volume.}\\ 

\section{Discussion}\label{sec:discution}
\begin{figure*}[ht]
    \centering
    \begin{adjustbox}{valign=t}
    \begin{subfigure}[t]{0.5\textwidth}
        \centering
        \includegraphics[width=\textwidth]{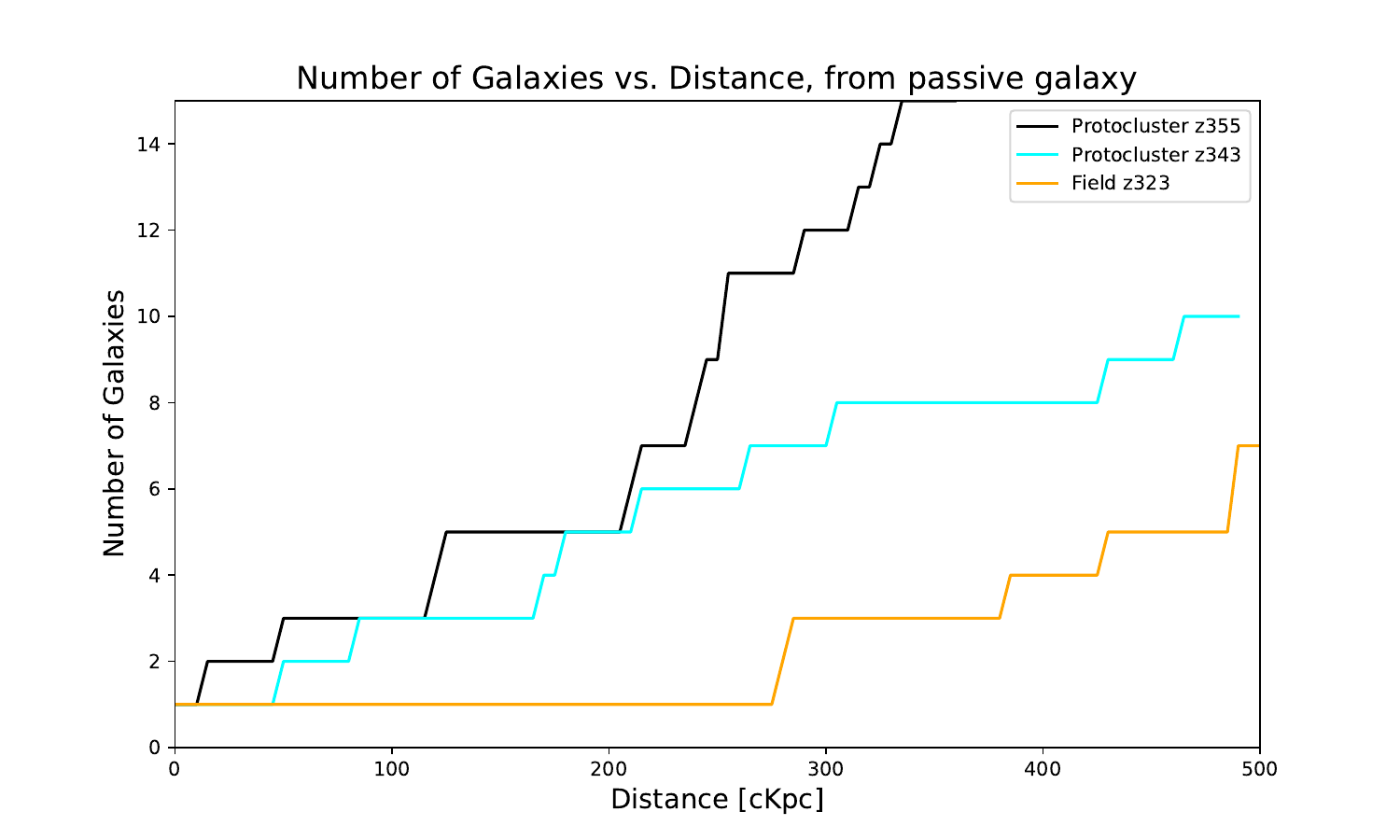}
        \caption{Overdensties}
        
        \label{fig:N_dist}
    \end{subfigure}
    \end{adjustbox}%
    ~ 
    \begin{adjustbox}{valign=t}
    \begin{subfigure}[t]{0.5\textwidth}
        \centering
        \includegraphics[width=\textwidth]{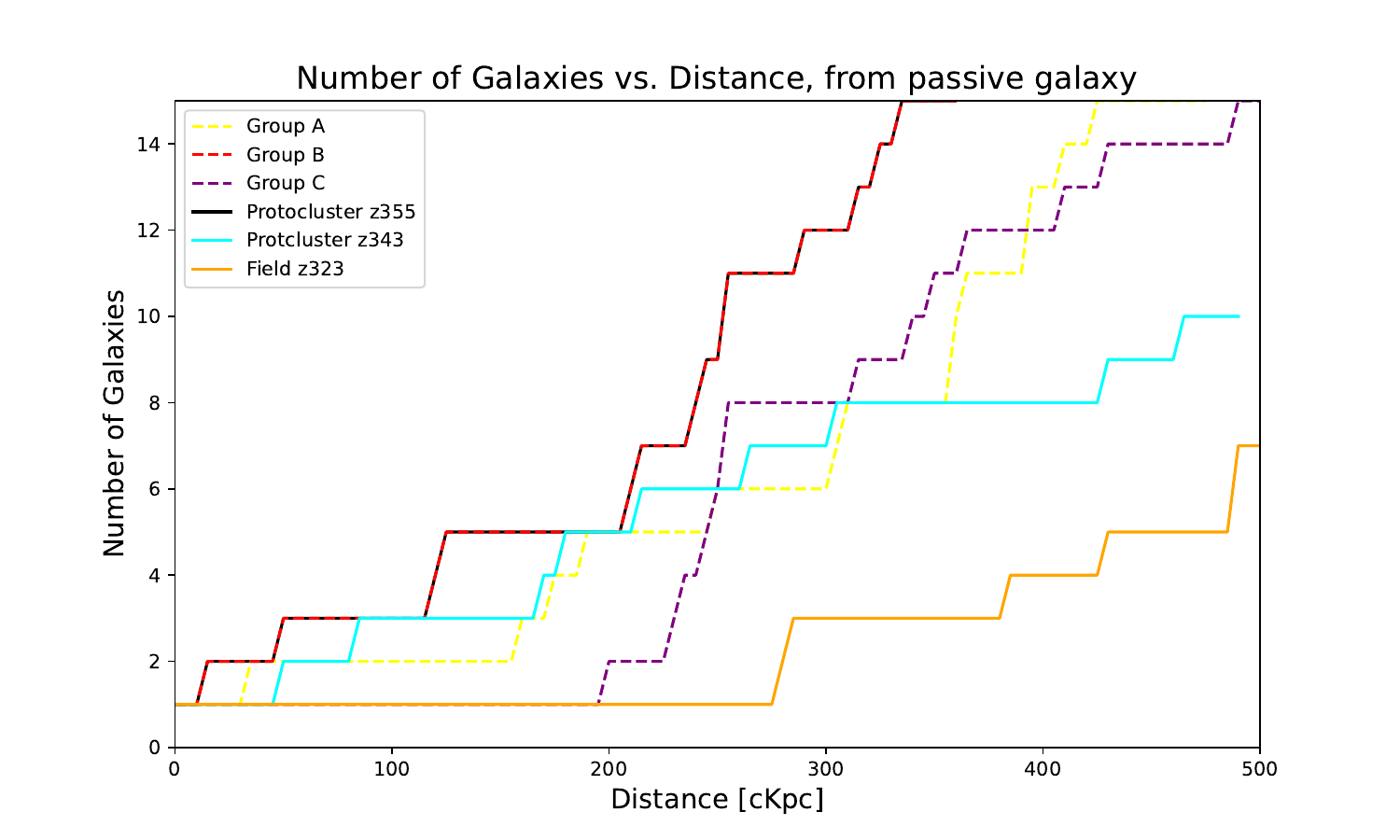}
        \caption{Group Differentiation}

        \label{fig:N_z355}
    \end{subfigure}
    \end{adjustbox}
    \caption{
    {Number of neighbors as a function of projected distance. The projected distance is calculated with respect to the passive galaxy hosted by the $z343$ over-density (cyan line), to the passive galaxy hosted by the $z355$ over-density (black line), and to the passive galaxy, from the literature, located in the field around the $z323$ over-density (orange line) in both the left and right panels.  
    The right panel also shows the number of neighbor as a function of projected distance from the passive galaxies, from the literature, identified in the subgroups of the $z355$ over-density (yellow and purple dashed lines)
    }
    }
    \label{fig:distances_all}

\end{figure*}

In this work, we investigated the physical properties of the members of the over-densities detected in the CDFS and UDS fields \citep{guaita2020} at \(3 < z < 4\).
By using the UVJ diagram and the criterion based on the low sSFR values, we identified two quiescent galaxies in two of the over-densities. 
The presence of only two quiescent galaxies in two over-densities raises questions about the expected frequency and characteristics of such galaxies within over-densities.
{Given the role of dense environments in accelerating star formation, we 
might 
expect star-forming and also some 
evolved 
galaxies in dense regions. 
However, the identification of 
only one passive galaxy in the $z355$ and one in the $z343$ over-density could suggest a more complex scenario, in which only specific conditions would lead to the quenching of star formation in over-densities at high redshift.\\ 
\\
Our study contributes
to the understanding of galaxy evolution within over-densities at \(3 < z < 4\),
by investigating the impact of both
internal and environmental factors.
By 
comparing our findings with current cosmological simulations, we offer 
some insights for the interpretation of our results. \\
\\
In examining the two over-densities containing quiescent galaxies, 
our findings indicate that they host AGN and that 
the passive galaxies are characterised by redder colours,  higher stellar masses, older stellar populations, 
and more compact morphologies that the other members. These properties are also found in the literature \citep[e.g.][]{Barro13, Barro17,vander14} for passive galaxies at similar redshifts. \\ 
\\
When compared to the simulations outlined in \citet{chiang13}, the masses we estimated for the dark matter halo of these over-densities 
(about $10^{13} M_\odot$) are consistent with those expected for proto-clusters at $3<z<4$, 
that could evolve into clusters of $3-10\times 10^{14} M_\odot$ 
at $z=0$. }This qualitative agreement reinforces the validity of our approach in identifying proto-clusters and highlights the similarities between our observed over-densities and the theoretical models. 
\\
Our analysis has led to a tentative estimation of the key epochs in the history of the passive galaxies, based on their star-formation timescale ($\tau$) and age,  
acknowledging the considerable uncertainty on these two parameters. 
Given the exponentially delayed form of the SFH model, for the passive galaxy in the $z355$ proto-cluster, we estimate the onset of formation at $z\sim6.20$, with the decreasing of SF already occurring at $z>4.3$. For the passive galaxy in the $z343$ proto-cluster, the formation time 
is traced back to $z\sim 4.8$, with a decrease in SF occurring at $z \gtrsim 3.5$.\\
\\
In Fig. \ref{fig:N_dist}, we compare the local environment around the passive galaxies in the proto-clusters with that around a passive galaxy, from the literature, located in the field associated to the $z323$ over-density {(this over-density does not host any passive galaxy).} We show the number of galaxies as a function of the projected distance from the passive galaxies, {where the 0 ckpc positions are calculated with respect to each passive galaxy.} 
Up to a distance of 280 ckpc,
our observations reveal a notable difference: more than two galaxies are found surrounding the passive galaxies hosted by the $z355$ and $z343$ over-densities, 
{while none are observed around the 
passive galaxy in the field. 
This may be expected since the proto-clusters, by definition, are regions denser than the field, but the passive galaxies are also located in the highest density peaks of the over-densities. 
The passive galaxy in the field is 
more isolated, reflecting the lower density and less clustered environment typical of field galaxies. 
This finding 
could highlight
the role of environment specifically in the evolution of the 
passive galaxies within our proto-clusters. 
While passive galaxies exist both in dense environments and in the field, 
the 
$z343$ and $z355$ proto-clusters may have encountered more interactions and been influenced by 
mechanisms, such as galaxy mergers or harassment \citep[e.g.][]{Moore1996}. These processes could have contributed to the quenching of star formation and to the morphological transformations of the quiescent galaxies.\\
\\
To quantify the environmental differences 
in the number of neighbors  
around the passive galaxies, 
we employed the Kolmogorov-Smirnov (KS) test. 
This statistical method allows us to compare the number of neighbors up to a spatial scale comparable to the projected size of the considered over-density.\\   
\\
Our findings reveal 
differences in the distributions of the number of neighbors around passive galaxies as a function of projected size. 
%
For the distributions around the two passive galaxies in the $z355$ and $z343$ over-densities, the KS test  
yields a KS statistic of 0.2 with 
a p-value of 0.182, 
which is not below the typical 0.05 value (confidence level of 95\%) and indicates that we cannot reject the null hypothesis that the curves in Fig. \ref{fig:distances_all} are drawn from the same distribution. 
A 
significant difference emerges 
when comparing each of the distributions around the two passive galaxies in the $z343$ and $z355$ over-densities with respect to that around the passive galaxy in the field. 
%
%
We calculated 
a KS statistics of 0.56 and 0.22, with p-values of \(1.5 \times 10^{-14}\) and \(6.5 \times 10^{-9}\), respectively. 
}\\
\\
{These results underscore the complex role of local environments in shaping galaxy evolution, particularly concerning passive galaxies. While the presence of passive galaxies both in over-densities and in the field 
might suggest that environment alone is not the main driver of the quenching of star formation, 
environmental factors could still have 
influenced the evolution of the passive galaxies in our over-densities. 
}\\
\subsection{Over-density $z355$}\label{sub:z355}
\begin{figure}[!htbp]
    \resizebox{\hsize}{!}{
    \includegraphics{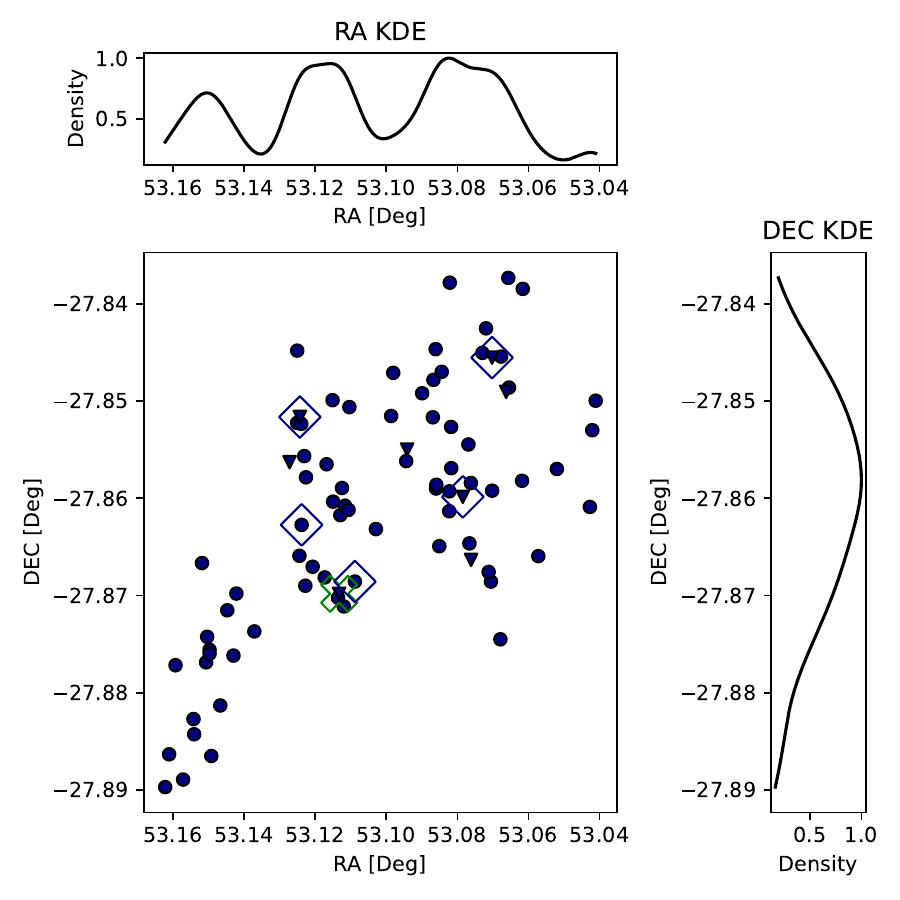}}
    \caption{Central panel: spatial distribution of the galaxies in the $z355$ over-density. 
    The galaxies are represented with different symbols based on their redshift, with circles for photometric redshifts and downward triangles for spectroscopic redshifts. Additionally, AGN are denoted with diamonds. 
    Top and right panels: kernel density estimations (KDE) 
    for the Dec. and for RA, respectively. 
    }
    \label{fig:kde_z355}
\end{figure}

\begin{figure}[!htbp]
    \resizebox{\hsize}{!}{
    \includegraphics{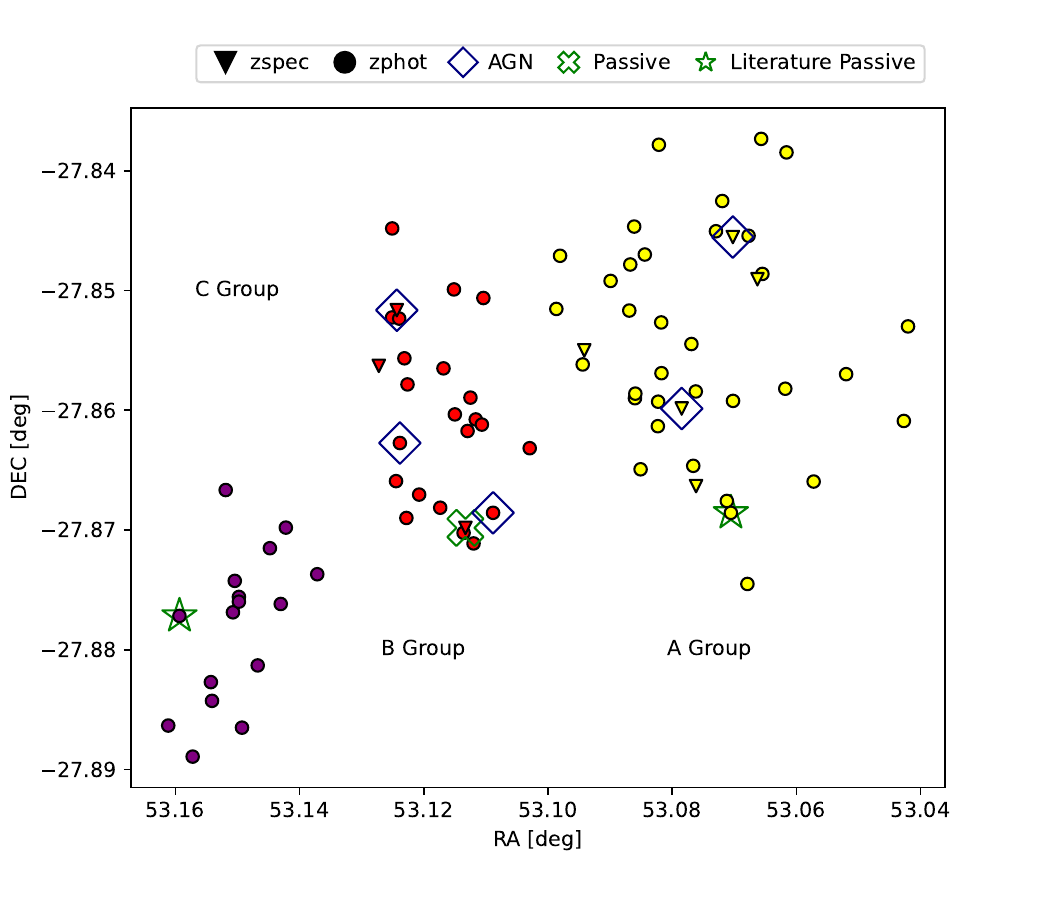}}
    \caption{
    {Distribution of the galaxies of the $z355$ over-density in the RA-Dec. plain. The galaxies are separated in three group with the kernel density estimation, as explained in the text and colour coded accordingly: 
    group A (yellow) at RA $\sim 53.08$, group B (red) at RA $\sim 53.12$, and group C (purple) at RA $\sim 53.15$. 
    The symbols follow the same scheme as in Fig. \ref{fig:kde_z355}. We also show two passive galaxies from the literature as green crosses. }
    }
    \label{fig:groups}
\end{figure}

\begin{figure}[!htbp]
    \resizebox{\hsize}{!}{
    \includegraphics{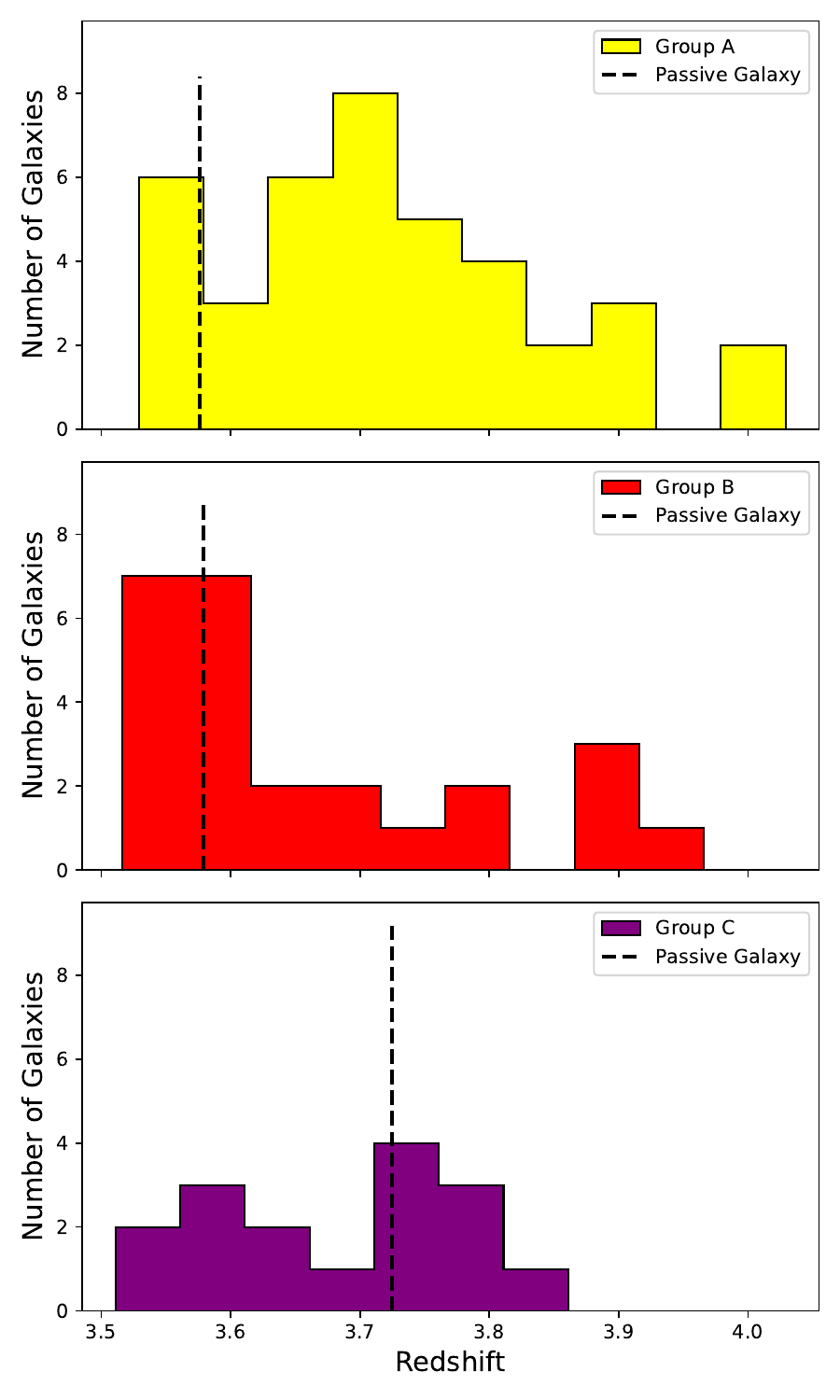}}
    \caption{{Redshift distributions 
    for each group in the \(z355\) over-density. 
    The top panel corresponds to group A, showcasing a peak at \(z\sim 3.7\); the middle panel 
    to group B, 
    with a peak at \(z\sim 3.55\); the bottom panel 
    to group C, with a peak at \(z\sim 3.75\). 
    The vertical lines 
    indicate the redshifts of the passive galaxy we identified in group B and the passive galaxies from the literature hosted by group A and B. 
    }
    }

    \label{fig:z_dist_groups}
\end{figure}

{The first over-density in which we identified a passive galaxy is that denoted with the $z355$ nomenclature in Table 1. In our investigation of the characteristics of the $z355$ over-density, a critical aspect 
is the spatial and redshift distribution of its members.
In Fig. \ref{fig:kde_z355}, we present the kernel density estimation \citep[KDE][]{Silverman} of the 
RA-Dec. diagram.} 
\\
{The KDE analysis leads us to subdivide the galaxies within the over-density into three 
groups, given their RA values (group A, B, and C). 
In Fig. \ref{fig:groups}, the galaxies in the three groups are shown in three different colours. The passive galaxy, we identified, 
is located in group B, the most spatially compact, in the middle of the over-density. 
Two additional passive galaxies from the literature are also shown, 
one is located in group A and the other in group C. 
The redshift distributions of the galaxies in each group is shown in Fig. \ref{fig:z_dist_groups}. 
The redshift range is of the order of 0.3-0.4 in all the groups. This range may be due to the fact that the majority of the galaxies in the $z355$ over-density has  photometric redshifts. However, our spectroscopically-confirmed passive galaxy is located at the peak of the redshift distribution of group B and this reinforces its association with group B. 
\\
Group A is the widest of the three in the right ascension, declination, and redshift space. Group C is like an elongated tale of the $z355$ over-density. 
The passive galaxy from the literature hosted by group A is at the lower limit of the redshift distribution; whereas that of group C is at the peak of the redshift distribution. The three groups and their wide redshift ranges 
may also indicate that the ovedensity as a whole is in an early stage of evolution that will settle in a more 
gravitationally bound structure later on. 
}
\\ 
\\
{Figure \ref{fig:distances_all} shows the number of galaxies as a function of projected distance  for the passive galaxy in group A and as well as that of group C. Their local environment is also richer than that of the passive galaxy in the field. The KS test quantifies the difference between the distributions around these two passive galaxies and the passive galaxy in the field. We calculated a KS statistics of 0.38 and 0.47, with p values of $2.5 \times 10^{-14}$ and $6.5 \times  10^{-9}$, respectively for the passive galaxy in groups A and C.}\\
\\
%
{The peak of the redshift distribution of group B is at 
\(z\sim 3.55\), that of group A is at  \(z\sim 3.7\).} 
\citet{shah2023} identified six massive protostructures within the redshift range of \(2.5 < z < 4.5\) in ECDFS, using public and proprietary spectroscopic data in addition to photometric redshifts. Our group A overlaps with a massive structure at $z \simeq 3.69$ (S5 structure from Table 1 in \citet{shah2023}). Our group B overlaps with a less massive structure at $z \simeq 3.58$ (Shah et al., private communication).\\
\\

\subsection{The $z343$ over-density}\label{sub:z343}
{The second over-density in which we identified a passive galaxy is the one denoted with the $z343$ nomenclature in Table 1. As we can see in Fig. B1, 
the redshift range of the $z343$ over-density is of the order of 0.2, narrower than that of the $z355$ over-density, and it is dominated by spectroscopic redshifts. 
However, this over-density is composed by much fewer members and so the passive galaxy is one out of only ten sources. 
%
%
The redshift spread in an over-density usually reflects the velocity dispersion of the member galaxies, that is influenced by the cluster's dynamical state and mass distribution. Therefore, 
the narrower redshift 
range
could be related to the availability of spectroscopic redshifts, but could also imply a more dynamically mature structure.\\ 
\\
In Fig. \ref{fig:N_dist} and Fig. \ref{fig:N_z355}, we present the number of galaxies as a function of projected distance from the passive galaxy in the $z343$ over-density with a cyan line. The local environment around this passive galaxy is also denser than that around the passive galaxy in the field. 
For this passive galaxy, it is also possible that
environmental factors are playing a role in the halt of its star formation. 
If the environment 
was playing a major role in the quenching in general, 
the passive galaxy in the field 
may have potentially undergone and survived past environmental interactions; for this reason, it may be currently isolated. However, this scenario cannot be proven with our data.}

\subsection{Comparison with previous studies}
\begin{figure}[!htbp]
    \resizebox{\hsize}{!}{
    \includegraphics{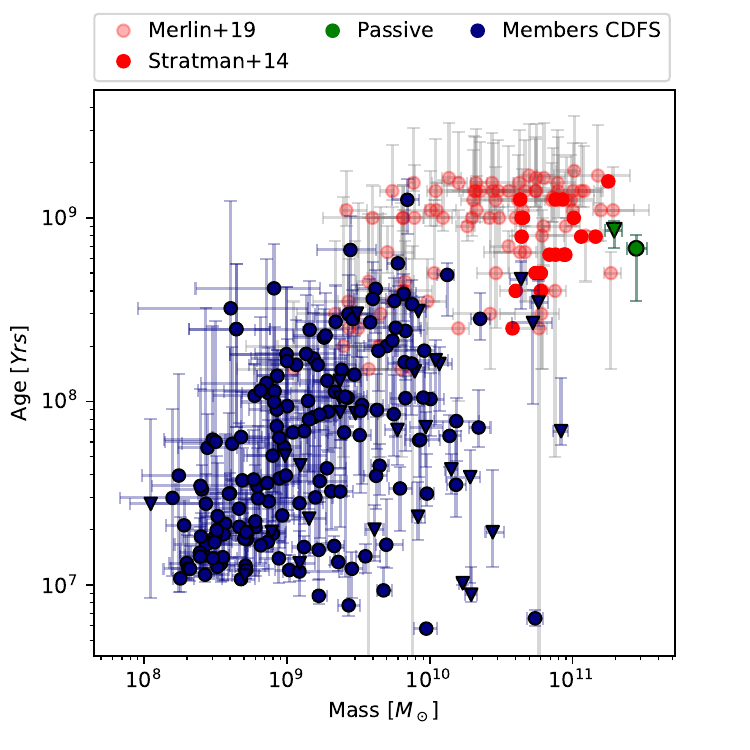}}
    \caption{Age as a function of stellar mass. Navy points represent the members of the over-densities in CDFS, with circle shapes indicating galaxies with photometric redshifts 
    and triangle-down shapes for those 
    with spectroscopic redshifts 
    Green points denote our identified passive galaxies. Light-red points correspond to passive galaxies from \citet{merlin19}, {at $3.01<z<3.9$}. Dark-red points represent passive galaxies from \citet{straat14}, {at $3.46<z<4.05$.}}

    \label{fig:literature_mass}
\end{figure}

{To contextualize 
the properties of the passive galaxies in over-densities, we compare with catalogs of passive galaxies from the literature. 
Specifically, we refer to 
the works of \citet{merlin19} and \citet{straat14}, which employ a combination of photometric analysis and 
SED fitting to identify and classify passive galaxies based on their age, stellar mass, and SFR.\\
\\
In Fig. \ref{fig:literature_mass}, we compare the age and stellar mass of the members of our over-densities in CDFS and the passive galaxies from the literature. Star-forming galaxies are the most common members of our over-densities and they are typically younger and less massive than passive galaxies. 
Their star-forming nature indicates that they are still undergoing an active growth that will increase their mass. 
Some of the passive galaxies from the literature overlap in age and mass with our star-forming members, indicating that passive galaxies in general are not necessarily massive. 
The ages and the masses of 
the passive galaxies we identified in our over-densities are consistent with the values found for the 
passive galaxies reported 
in the literature. 
Additionally, they are among the most massive passive galaxies, making the largest gap in mass between star-forming and passive galaxies in over-densities. This may be related to the accelerated phenomena that favor the mass growth and then the star-formation quenching in the over-densities. These phenomena could be related to environmental interactions. For instance, the local density of passive galaxies in over-densities may be higher than that of passive galaxies in the field. Other factors may include accelerated gas accretion and AGN feedback, as the over-densities with AGNs in our sample are the only ones that contain passive galaxies. However, AGN feedback can also affect galaxies in a variety of environments and in a variety of time scales \citep[e.g.][]{mao22,KurinchiVendhan2023}.
}\\

\subsection{Comparison with a cosmological simulation}
\begin{figure}[!h]
    \resizebox{\hsize}{!}{\includegraphics{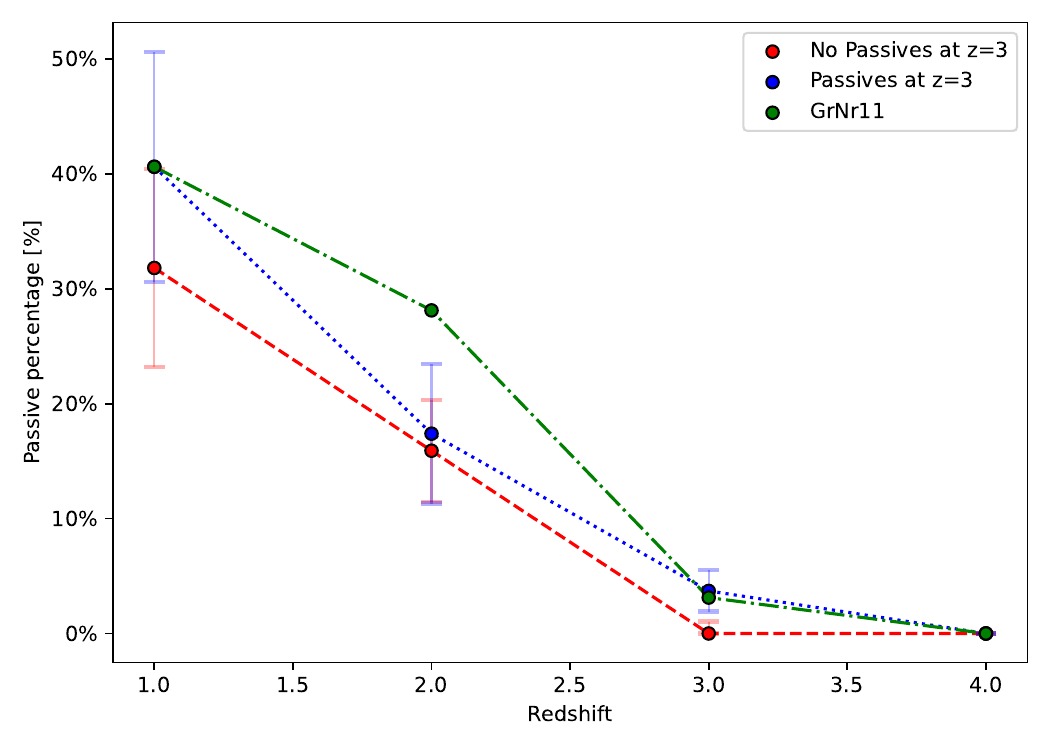}}
    \caption{Fraction of passive galaxies as a function of redshift in the simulated proto-clusters. 
    Blue points correspond to the median values calculated for the 15 proto-clusters hosting passive galaxies at $z=3$; red points for the 11 proto-clusters only hosting star-forming galaxies at $z=3$. The green points correspond to the fractions of the GrNr11 simulated proto-cluster. This is a representative proto-cluster for us, because it contains a passive galaxy as well as star-forming galaxies with sSFR larger than 10$^{-8}$ yr$^{-1}$, as in our observed ovdendensitis. 
    }
    
    \label{fig:fraction_passive}
    
\end{figure}

{We utilised the Illustris TNG300 simulation \citep{nelson19} to investigate the evolutionary path 
of the passive galaxies. We identified in the $z343$ and $z355$ over-densities, and also to investigate the evolution of the over-density hosting passive galaxies at $z\sim3$.  
In particular, we used the TNG300 box that has a side length of about 300 Mpc.
Andrews et al. (in prep.) selected the 30 largest clusters in TNG300 at 
$z=0$. Centering around the barycenter of each of these clusters, they defined a 60 cMpc$^3$ volume in the $z = 3$ snapshot and identified the proto-clusters, progenitors of the 30 clusters. 
We consider the galaxies, members of these proto-clusters, according to their Subhalo IDs. For each of these galaxy members, we recovered their SFR and stellar mass 
information from $z=8$ to $z=0$ \citep{pillepich18,naiman18}. We also verify that the 
dark matter halos of the simulated proto-clusters match the mass values of the $z343$ and $z355$ over-densities.\\
Among the simulated structures, we searched for the most representative ones for our observations. Therefore, we chose the proto-clusters at $z=3$ hosting massive passive galaxies, characterised by stellar masses larger than $10^{11}$ M$_{\odot}$ and sSFR $\lesssim -10$. 
We identify 15 proto-clusters hosting either one or two massive passive galaxies. An example is the GrNr11 proto-cluster that contains a passive galaxy with a mass of $6\times 10^{11}$ M$_{\odot}$ and also contains star-forming galaxies with sSFR larger than $10^{-8}$ yr$^{-1}$, as we have in our observed over-densities.
Using the method in \citet{burg14} for the dark matter halo mass estimation, we determined that GrNr11 
possesses a dark matter halo mass of 
$3 \times 10^{13} M_{\odot}$.}
\\
\\
To elucidate the evolutionary trajectories of the passive galaxies within the 15 proto-clusters across different redshifts, we considered
the sSFR evolution and stellar mass growth. 
{
The median sSFR of the passive galaxies was larger than 10$^{-8}$ yr$^{-1}$ at $z=6$ and the median mass growth rate was about 60\% from $z=6$ to $z=4$ and about 40\% from $z=4$ to $z=3$. In total, the typical galaxy had a mass growth rate of 96\% from $z=6$ to $z=3$, which means that, at $z=6$, the galaxies had 4\% of the mass built till $z=3$. However, 20\% of the proto-clusters have a passive galaxy, which  at $z=6$, had already accreted 10-20\% of the mass at $z=3$. These galaxies had SFRs greater than 100 M$_{\odot}$ yr$^{-1}$ at $z=8$ and sSFR $>$10$^{-8}$ yr$^{-1}$ at $z\gtrsim6$, in agreement with the notion that star-formation activity may be the main driver of mass growth. 
The passive galaxy in the proto-cluster GrNr11, 
for instance, exhibits a 
sSFR of the order of 10$^{-8}$ yr$^{-1}$ and SFR = 279 M$_\odot$ yr$^{-1}$ at $z=6$, when its mass was 10\% of the mass accumulated till $z=3$. 
At $z\leq2$, the median behaviour of the passive galaxies is that they remain passive. 
%
\citet{KurinchiVendhan2023} studied the evolution of quiescent galaxies at $z\sim3$ in the TNG simulation as opposed to the evolution of star-forming galaxies. They found that, before becoming passive, quiescent galaxies were typically surrounded by a larger number of neighbors within 1 cMpc than star-forming galaxies. In these denser regions, they experienced a higher gas accretion rate from $z=8$ on, that could favor a higher star-formation activity and also a faster growth of their central black hole mass. As a consequence, these galaxies have been affected by AGN feedback, responsible for shutting down their star formation, early on in their evolution, more than star-forming galaxies.  
These findings are in agreement with the evolutionary paths we see for the massive passive galaxies in the 15 proto-clusters. 
\\
\\
As a control sample, we also identify 11 proto-clusters at $z=3$ that only host star-forming galaxies.}
To study the evolution of the over-densities hosting passive galaxies and of those containing only star-forming galaxies, we focus on the quiescent fraction which is the number of quiescent galaxies 
divided by the total number of members as a function of redshift (Fig. \ref{fig:fraction_passive}). 
\\
{At $z=4$, the quiescent fraction is close to 0\% for all the proto-clusters. At $z=3$, the median value is 3.7\% for the 15 proto-clusters, hosting passive galaxies, and it becomes more then 40\% at $z=1$. The median value for the 11 proto-clusters, without passive galaxies at $z=3$, is estimated to be 30\% (10\% lower than the previous case) at $z=1$.  
One possibility is that for the proto-clusters that do not show a passive galaxy already at $z=3$, the gas accretion rate would increase at later epochs, for instance, producing a delayed start of the star formation activity and feedback phenomena.}
\\
For the $z355$ over-density, we calculate a quiescent fraction of 1.21\%, which increases up to 3.65 \% if we include 
the passive galaxies from the 
literature. These values agree with the median fraction obtained for the proto-clusters from the TNG300 simulation.
The $z343$ over-density showcases a higher passive galaxy fraction of 10\%.\\
\\

\section{Conclusions}
\label{sec:conclusions}
In this work, we studied the characteristics and the distribution of the galaxies within 13 overdensities, identified at $3<z<4$ in the CDFS and UDS field of the VANDELS survey \citep{guaita2020}. Our key findings can be summarised as follows:\\
\begin{itemize}

    \item {Among the 13 over-densities, only two are found to contain passive galaxies (Fig. \ref{fig:uvj_all}). 
    The two over-densities have masses compared to those of proto-clusters at $z\sim3$ \citep[eg.,][]{chiang13}.}\\

    
    \item {The passive galaxies, we identified in the two over-densities, were previously discovered in the literature \citep{straat14, merlin19}. They share physical properties with the old, massive passive galaxies at $3<z<4$ from the literature.
    They are redder, older, and more massive compared to other over-density members (Fig. \ref{fig:uv_Mass}). Also, they are characterised by compact morphologies (Fig. \ref{fig:Size_Mass}) when compared to the relations of size as a function of stellar mass for early-type and late-type galaxies at $z\sim 3$ \citep{vander12}.}\\
    

    
    \item 
    {The $z355$ over-density, which hosts a spectroscopically-confirmed quiescent galaxy at $z\simeq3.6$, can be separated in three groups according to the RA values of the member galaxies 
    (Fig. \ref{fig:kde_z355}).  
    The passive galaxy we have identified is locate in the highest density peak of the central group. Other two passive galaxies from the literature are located in the other two groups (Fig. \ref{fig:groups}). 
    The local density environment of the passive galaxies in the over-densities is estimated to be denser than that around a passive galaxy in the field (Fig. \ref{fig:distances_all}). This could indicate that the environment (with galaxy interaction and gas accretion) has played a role in the evolution of those passive galaxies.} \\ 
    
    \item {By cross-matching with catalogs of X-ray selected \citep{koce18, chandra} and emission-line (A. Bongiorno private communication) selected AGNs, we found AGN only in the over-densities that also host passive galaxies. 
    This suggest a potential link between AGN activity and the presence of passive galaxies in these dense environments. 
    If the environment is favoring interactions and gas accretion, the consequence would be an enhanced star-formation activity and also an enhanced black hole mass growth, that can be translated into AGN feedback. While AGN have been used to pinpoint over-dense regions \citep[e.g.][]{pentericci2000, kurk2004}, imprints of AGN feedback are also found in isolated galaxies in the field \citep[e.g.][]{Ito2022}.  
    Further investigation 
    will aim 
    at exploring the 
    connections 
    between AGN and their host galaxies in over-densities.}\\
    
    \item {In our investigation of the galaxy evolution within the observed over-densities, we utilised the IllustrisTNG (TNG300) simulation. Among the 30 simulated proto-clusters at $z=3$, we chose the most representative ones. We identified 15 proto-clusters with massive passive galaxies (M$_{*} \gtrsim 10^{11}$ M$_{\odot}$ and sSFR $\lesssim -10$), in addition to star-forming galaxies.
    The median sSFR of the passive galaxies was larger than 10$^{-8}$ yr$^{-1}$ at $z=6$ and the median mass growth rate was  96\% from $z=6$ to $z=3$. However, 20\% of the proto-clusters have a passive galaxy that, at $z=6$, had already accreted 10-20\% of the mass at $z=3$. These galaxies had SFRs larger than 100 M$_{\odot}$ yr$^{-1}$ at $z=8$ and sSFR $>$10$^{-8}$ yr$^{-1}$ at $z\gtrsim6$. 
    They could have experienced a higher gas accretion rate from $z=8$ on, which could favor a higher star-formation activity and also a faster growth of their central black hole mass. As a consequence, these galaxies have been affected by AGN feedback, responsible for shutting down their star formation, early on in their evolution, more than star-forming galaxies \citep{KurinchiVendhan2023}.  
    %
    As a control sample, we also considered 11 proto-clusters at $z=3$ that only host star-forming galaxies.
    The median value of the quiescent fraction for the 11 proto-clusters, without passive galaxies at $z=3$, is estimated to be 10\% lower than that for the 15 proto-clusters with passive galaxies (Fig. 
\ref{fig:fraction_passive}).}  
    %
\end{itemize}
{In conclusion, in our study, we found massive passive galaxies in two over-densities at $z\sim3$. In these over-densities we also found AGN. The passive galaxies are located in a dense local environment, that could have played a role in their evolution. In particular, galaxy interaction and gas accretion could have favoured star formation activity early on as well as AGN feedback. This scenario can explain the origin of our passive galaxies at $z\sim3$. }
%
%
%
%
\\
%
Given the complexity of the processes that happen in dense environments, 
further investigations are needed to disentangle the relative contributions of different quenching mechanisms. This includes interaction, gas accretion, and AGN feedback 
in shaping the evolution of galaxies in the early universe.

\section{Data availability}\label{data_av}

The supplementary figures related to this article are available on Zenodo (\url{https://zenodo.org/records/13909612}). These figures include UVJ diagrams of galaxy members in proto-clusters and histograms of redshift ranges for the identified over-densities in the CDFS and UDS fields. These datasets complement the results presented in the article, offering additional visual insights into the galaxy populations and redshift distributions.

\begin{acknowledgements}
    ME and LG acknowledge the FONDECYT regular project number 1230591 for financial support. Also, we thank Alain Andrade, Ian Baeza, Benjamin Jara Bravo, Sarath Satheesh Sheeba, Benjamin Forrest, Ekta A Shah, Brian Lemaux, and Olga Cucciati for useful discussions.
    ME, LG, and RD  gratefully acknowledges support by the ANID BASAL project FB210003.
\end{acknowledgements}

\bibliographystyle{aa} 
\bibliography{biblio} 

\begin{appendix}
\onecolumn
\section{Additional figures}

\begin{figure*}[!htbp]
    \centering
    \includegraphics[width=\textwidth]{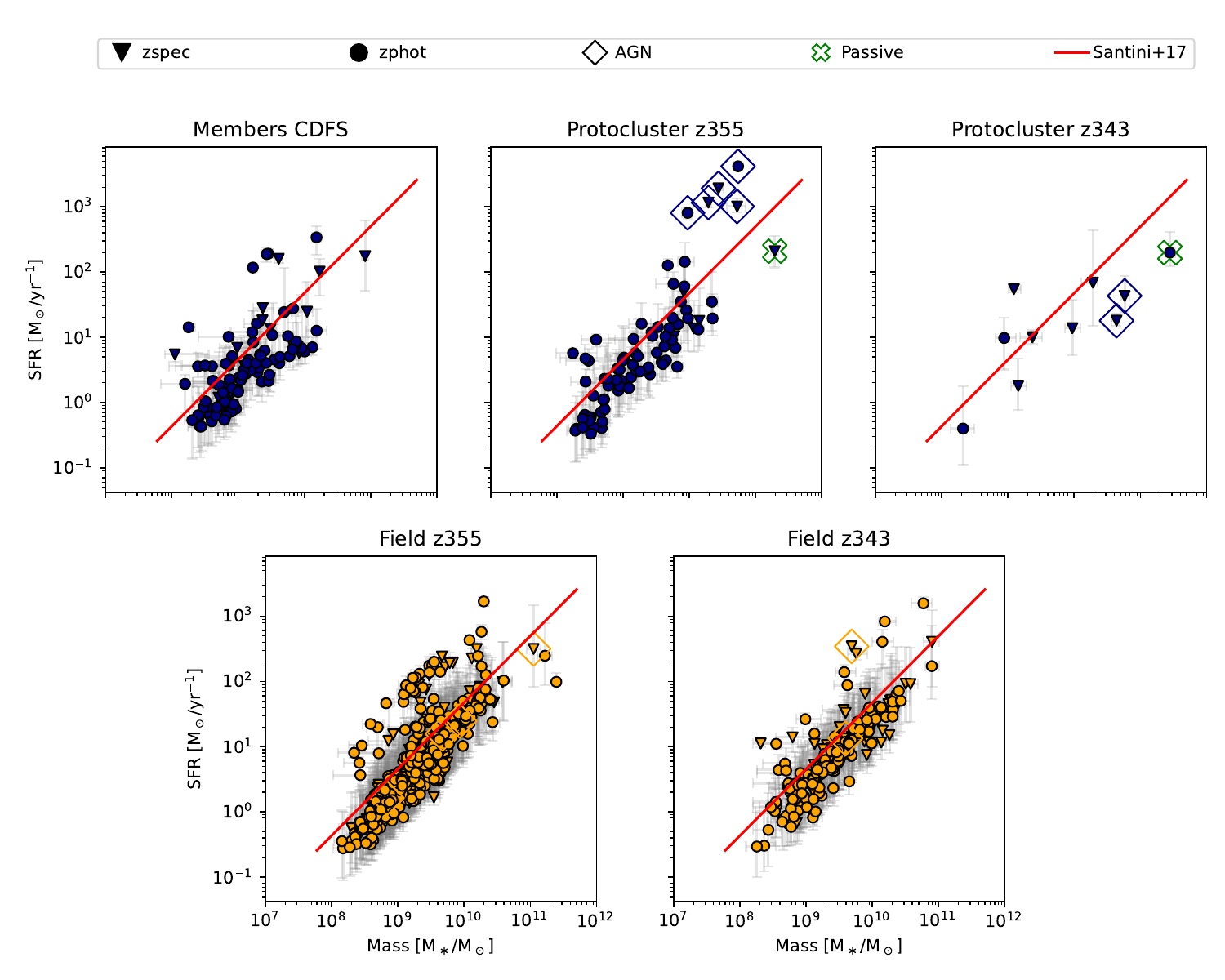}
    \caption{ {Star formation rate plotted against stellar mass for galaxies in different environments. Top-left: Galaxies that are members of the over-densities identified in CDFS excluding the two that host passive galaxies. Top-centre: members of the proto-cluster at z=3.55. Upper right: members of the proto-cluster at z=3.43. Bottom-left: field galaxies associated with the proto-cluster at z=3.55. Bottom-right: Field galaxies associated with the proto-cluster at z=3.43.
    The galaxy categories represented by the different symbols are the same as in Fig. \ref{fig:uvj_all}. 
    }
    }
    \label{fig:SFR_Mass}
\end{figure*}

\begin{figure*}[!htbp]
    \centering
    \includegraphics[width=\textwidth]{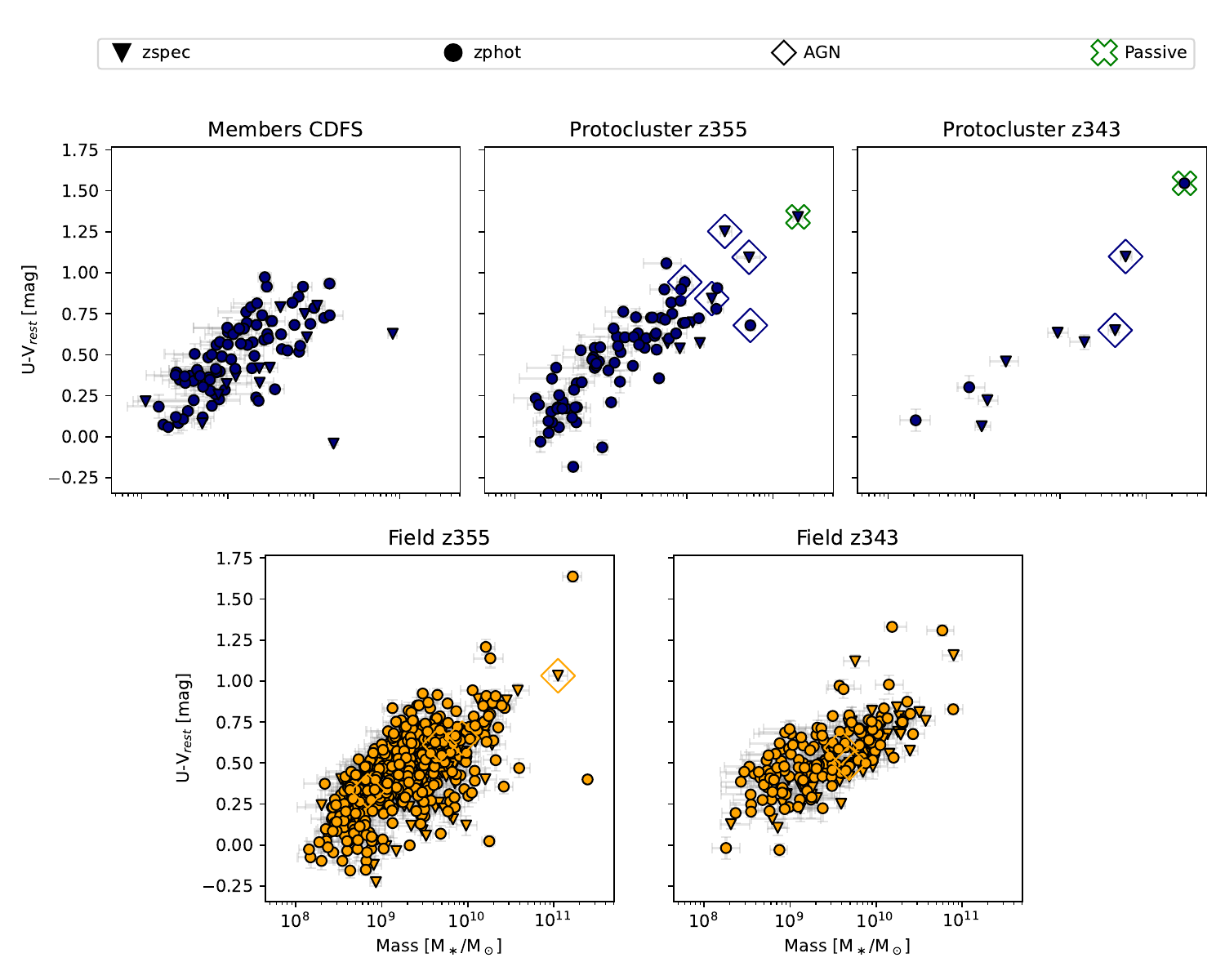}
    \caption{Rest-frame U-V as a function of stellar mass for the galaxies plotted in Figs. \ref{fig:uvj_all} and \ref{fig:SFR_Mass} with the same symbols.}
    \label{fig:uv_Mass}
\end{figure*}

\begin{figure*}[!htbp]
    \centering
    \includegraphics[width=\textwidth]{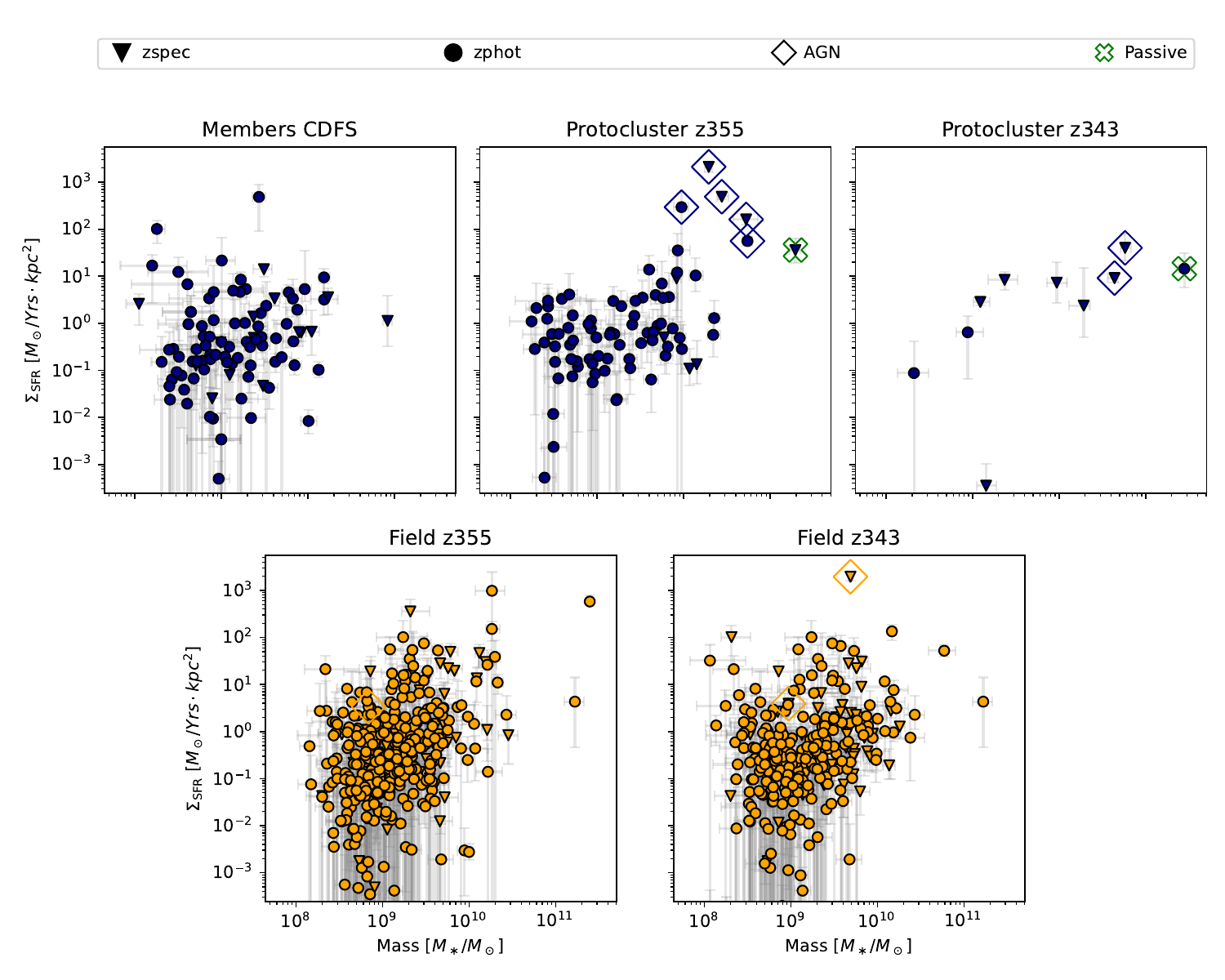}
    \caption{Star formation rate surface density as a function of stellar mass for the galaxies plotted in Figs. \ref{fig:uvj_all}, \ref{fig:SFR_Mass}, and \ref{fig:uv_Mass} with the same symbols.}
    \label{fig:SigmaSFR_Mass}
\end{figure*}

\begin{figure*}[!htbp]
    \centering
    \includegraphics[width=\textwidth]{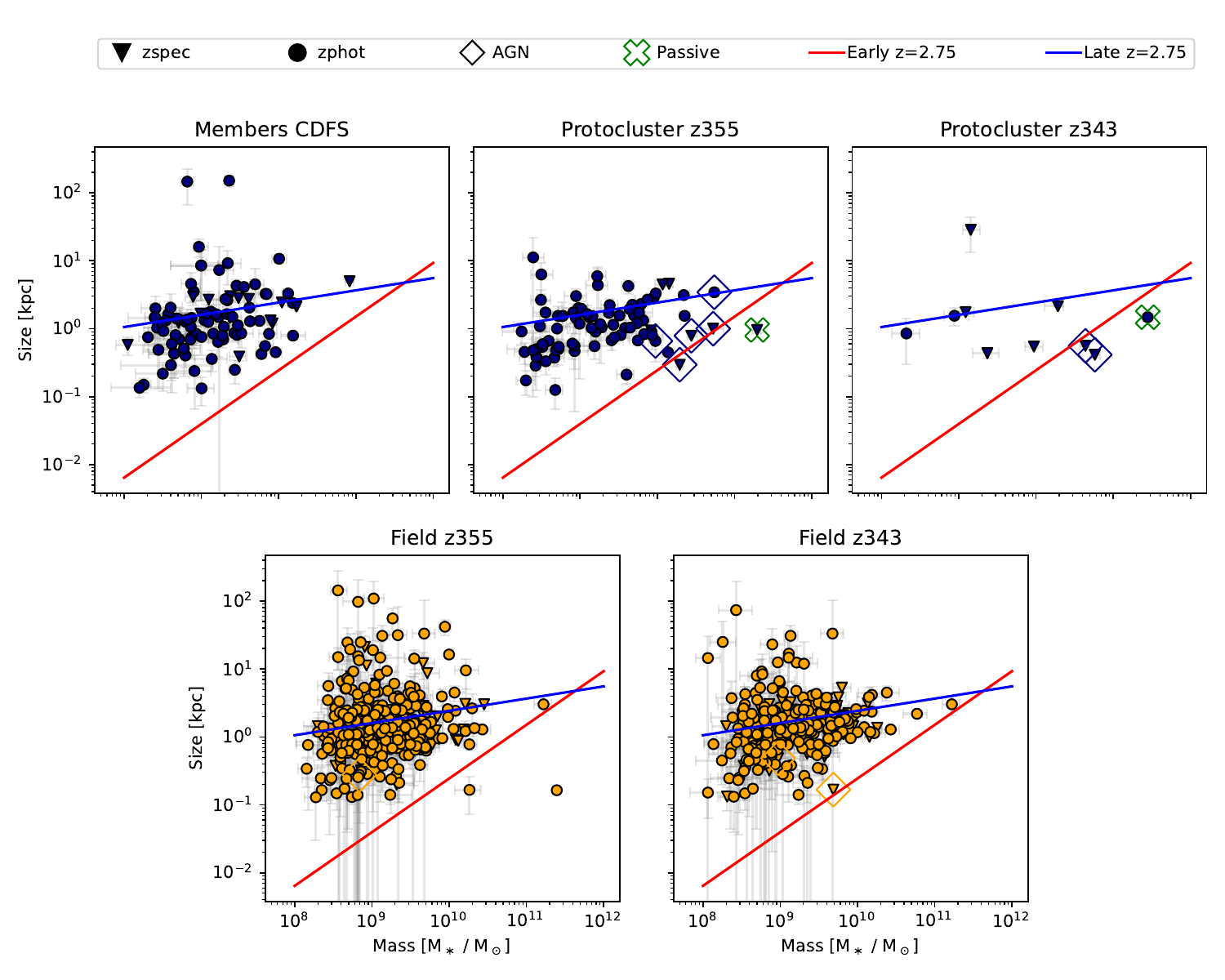}
    \caption{Size, parameterised by half light radius, as a function of  stellar mass
    for the galaxies plotted in Figs. \ref{fig:uvj_all}, \ref{fig:SFR_Mass}, \ref{fig:uv_Mass}, and \ref{fig:SigmaSFR_Mass} with the same symbols. The lines in the diagrams represent the relation found in \citet{vander14} for galaxies at $z=2.75$.}
    \label{fig:Size_Mass}
\end{figure*}

\end{appendix}

\end{document}